\documentclass[journal]{IEEEtran}
\ifCLASSINFOpdf
\else
\fi
\usepackage{cite}
\usepackage{amsmath,amssymb,amsfonts}
\usepackage{algorithmic}
\usepackage{graphicx}
\graphicspath{{Figures/}}
\usepackage{textcomp}
\usepackage{xcolor}
\usepackage{url}
\usepackage[utf8]{inputenc}
\usepackage[normalem]{ulem}
\usepackage{booktabs}

\begin{document}
%
\title{dAJC: A 2.02mW 50Mbps Direct Analog to MJPEG Converter for Video Sensor Node using Low-Noise Switched Capacitor MAC-Quantizer with Auto-Calibration and Sparsity-Aware ADC}
%
%
%

\author{\IEEEauthorblockN{,  and Shreyas Sen}
\IEEEauthorblockA{Elmore Family School of ECE, Purdue University, West Lafayette, USA-47906. email:\{gbarik, gauravk, bchatte, shreyas\}@purdue.edu}}

\author{Gourab Barik, \IEEEmembership{Graduate~Student~Member,~IEEE}, Gaurav~Kumar~K,\IEEEmembership{ Graduate~Student~Member,~IEEE},  Baibhab~Chatterjee, \IEEEmembership{Member,~IEEE}, Shovan~Maity, \IEEEmembership{Member,~IEEE}, Sumon~Bose, \IEEEmembership{Member,~IEEE}, and~Shreyas~Sen,~\IEEEmembership{Senior~Member,~IEEE}

\thanks{}

\thanks{This work was supported by Quasistatics Inc. \textit{(Corresponding author: Shreyas Sen).}}

\thanks{The following authors are with the Elmore Family School of Electrical and Computer Engineering, Purdue University, West Lafayette, IN 47907 USA e-mail: \{gbarik,gauravk,shreyas\}@purdue.edu. The following author is with the Electrical and Computer Engineering, University of Florida, Gainsville, FL 32611 e-mail: \ {chatterjee.b}@ufl.edu. The following authors are with  Quasistatics Incorporation. West Lafayette, IN 47906. e-mail: \{shovan, sumon.bose\}@10xar.com

}

 \thanks{}
 }

%
%

\markboth{}
{Barik \MakeLowercase{\textit{et al.}}: dAJC}
%



\maketitle

\begin{abstract}
\textcolor{black}{} With the advancement in the field of the Internet of Things(IoT) and Internet of Bodies(IoB), video camera applications using Video Sensor Nodes(VSNs) have gained importance in the field of autonomous driving, health monitoring, robot control, and security camera applications. However, these applications typically involve high data rates due to the transmission of high-resolution video signals, resulting from high data volume generated from the analog-to-digital converters (ADCs). This significant data deluge poses processing and storage overheads, exacerbating the problem. 
To address this challenge, we propose a low-power solution aimed at reducing the power consumption in Video Sensor Nodes (VSNs) by shifting the computation from the digital domain to the inherently energy-efficient analog domain.
Unlike standard architectures where computation and processing are typically performed in digital signal processing (DSP) blocks after the ADCs, our approach eliminates the need for such blocks. Instead, we leverage a switched capacitor-based computation unit in the analog domain, resulting in a reduction in power consumption. We achieve a $\sim4X$ reduction in power consumption compared to digital implementations. 
Furthermore, we employ a sparsity-aware ADC, which is enabled only for significant compressed samples that contribute to a small fraction ($\le5\%$) of the total captured analog samples, we achieve a $\sim20X$ lower ADC conversion energy without any considerable degradation, contributing to the overall energy savings in the system.
\end{abstract}


%
\IEEEpeerreviewmaketitle

\section{Introduction}

\IEEEPARstart 
Emerging applications in the IoT field, such as robot control, health monitoring, autonomous driving, and security cameras, necessitate the capture of high-resolution video data \cite{kondo2003ultra}. Traditionally, this captured analog video data is digitized before further processing, consuming a significant amount of ADC energy during the conversion process while generating enormous data volumes. 
Although standard wireless communication modalities like Bluetooth have low data rates and cannot support the transfer of this data deluge, the 2.4 GHz and 5 GHz WiFi systems \cite{6353239} offer sufficiently higher data rates, such as 600 Mbps and 1.3 Gbps at the cost of high pJ/bit for transmission.
However, this poses a challenge due to the high power consumption required for computation and communication at these rates, ultimately creating a technical bottleneck for the widespread adoption of such Video Sensor Nodes(VSNs).
The current demand for these devices is twofold: they must handle high data volumes while also being compact in form factor. Additionally, they need to efficiently communicate captured video data with nearby hubs, highlighting the importance of efficient communication and computing mechanisms in these devices.

\begin{figure}[t]
   \centering
    \includegraphics[width=0.48\textwidth]{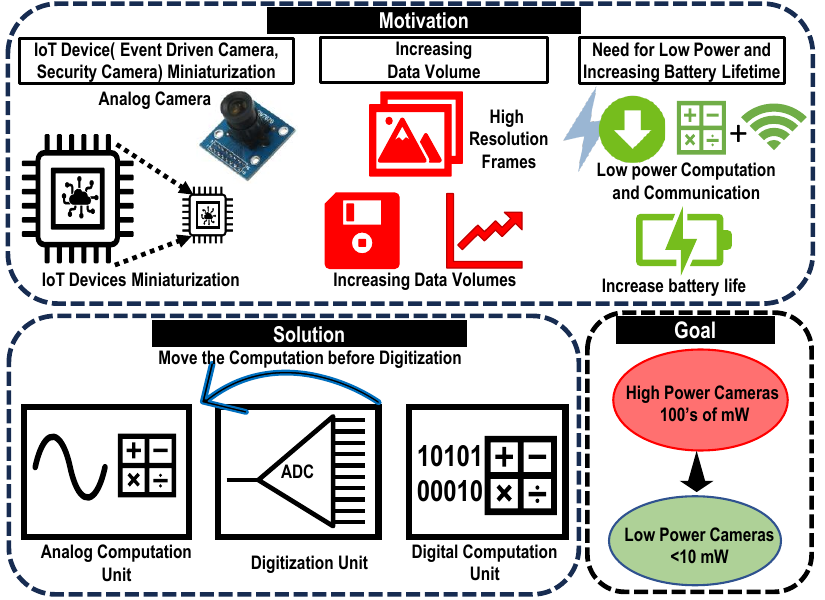}
    \caption{Miniaturization of IoT devices and implementation of cameras in smaller form factor along with the increase in the data volume due to increasing video frame size requires low power processing cameras to remove the need for frequent battery replacement and extend battery lifetime}
    \label{fig:LP_CAM_motivation}
    \vspace*{-0.25in}
\end{figure}

\begin{figure*}[t]
   \centering
    \includegraphics[width=0.85\textwidth]{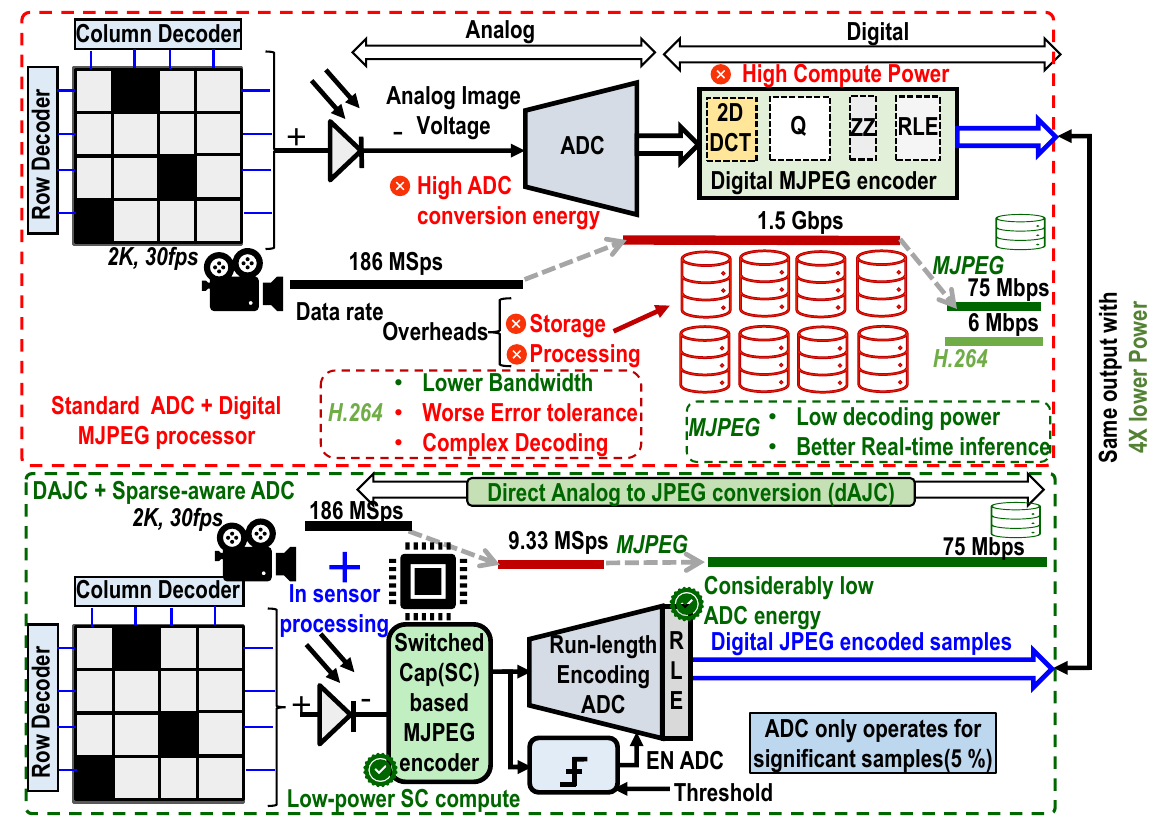}
    \caption{Comparison between Standard MJPEG compression approaches against our approach. dAJC has the advantage of compression of the information in the analog domain before digitizing, which prevents the increase in the intermediate amount of data(reducing intermediate storage and processing elements). The addition of the sparsity-aware ADC reduces the ADC energy since the ADC is turned ON only for significant samples.}
    \label{fig:Motivation_intro}
\end{figure*}
The demand for smaller form factors in IoT devices often limits the size of the battery that can be integrated, necessitating frequent recharging or battery replacement for high-power-consuming systems. Despite efforts to increase battery life through power-gating techniques and sub-threshold circuit design, higher video resolutions and frame rates escalate analog input data rates, thereby exacerbating power consumption and the need for battery maintenance. Hence a solution is needed to handle the incoming data of 100's of Mbps, without having to digitize all of the data to lower the overall power consumption of the video sensor nodes and remove the need for any storage overheads as shown in Fig. \ref{fig:LP_CAM_motivation}. 

To address these challenges, our solution leverages analog domain compression using a switched-capacitor-based DCT unit. By compressing the analog input data in the analog, this approach reduces overall data volume removing the need for any DSP units post ADC. Moreover, implementing a sparsity-aware ADC further enhances power efficiency by digitizing only significant samples, exploiting the inherent sparsity in image data. This integrated approach minimizes power consumption, alleviates storage overheads, and extends battery life, ensuring sustained performance in high-data-rate IoT environments.



In a typical IoT system, transmitting sensor data wirelessly to a processing hub for storage demands careful consideration of communication power to optimize energy usage. The power consumption for wireless transmission, as indicated by Eq. \ref{comm_power}, is directly proportional to the number of transmitted bits and the energy efficiency of the communication method \cite{9249386}. For instance, conventional WiFi communication operates at an energy efficiency of 1-10 nJ/bit.
\begin{equation} \label{comm_power}
    \text{Tx power}  = \text{No. of bits} \times \text{Energy Eff.}
\end{equation}
Considering a scenario where a 2K RGB video at 30fps with 8-bit digitization results in a substantial data volume of 1.5 Gbps. Transmitting this data via WiFi would consume 1.5-15 W of power according to Eq. \ref{comm_power}. However, employing compression techniques like MJPEG \cite{okada1997single} can significantly reduce the data rate to 600 Mbps, thereby lowering the power consumption to 0.6-6W during WiFi transmission. This demonstrates a notable decrease in communication power through standard video compression.

Moreover, adopting more energy-efficient communication modes like Human Body Communication (HBC) or Ultra-Wide Band (UWB) offers additional power savings. HBC, suitable for body-worn devices, and UWB, ideal for personal area networks (PAN), boast energy efficiencies of 10-100 pJ/bit, making them 100 times more efficient than standard WiFi. While HBC\cite{9271808} may not support Gbps data rates, its suitability for transmitting compressed data in the Mbps range ensures reduced digitized bits, resulting in lower communication payload and energy consumption.

As shown in Fig. \ref{fig:Motivation_intro}, a Video Sensor node typically employs a camera that captures 
the analog image data from the image sensor in the form of analog voltage samples. It is followed by an ADC that converts the analog samples to digital bits followed by compression in the digital domain(using DSP) before communicating the bits, which consumes 50mW-100mW of power. To get an estimate of the overall data volume, a sample calculation has been shown for a gray-scale image. A gray-scale video of 2K resolution generates data at the rate of 186 MSps which when digitized using 8-bit ADC generates a huge amount of digital bits at 1.5Gbps leading to high power consumption in the digital signal processor to compress the digital bits by 20$\times$ to 75 Mbps (for MJPEG). In this process, the intermediate data generated requires higher storage and processing overheads. However, our approach of dAJC with Sparsity-Aware ADC does in-sensor processing of the received analog samples. Here, the analog samples received at 186 MSps were compressed in the analog domain to 9.33 MSps before digitization. Since the compression is done in the analog domain using a switched capacitor-based MJPEG encoder, the overall power required is significantly lower than the digital compressing methods. The compressed analog samples have sparsity in the data, that is they contain very few significant samples and all other samples are small and insignificant. The sparsity-aware ADC takes advantage of this sparsity to digitize only the significant samples and the remaining time the ADC is turned OFF, hence reducing the overall ADC conversion energy. 

The implemented system works for JPEG compression of analog images (Image pixels represented in the form of analog voltage)\cite{kumar202365}. This system can be extended to videos that rely on intraframe dependencies like MJPEG. Other compression methods such as H.264 \cite{wiegand2003overview} utilize the inter-frame redundancies and offer a higher degree of compression. However, H.264 has a complex decoding scheme and has poor error tolerance, therefore failing to work for applications that are latency-sensitive and suffer from error-prone communication. Compared to that MJPEG benefits from per-frame compression and has a better error tolerance and a less complex decoding scheme which makes it ideal for applications requiring real-time inference and decision-making. Lastly, MJPEG enables a simpler hardware design compared to the H.264 compression technique.

\subsection{Contributions of our work}
\begin{itemize}
    \item This work, \textbf{d-AJC} presents the first end-to-end implementation of MJPEG compression in the mixed-signal domain. 
    \item We achieved $\sim4$$\times$ lower power consumption than its digital counterparts, illustrating the benefit of energy-efficient analog computing.
    \item The switched capacitor-based low-power 2D-DCT core consumes $\sim$12$\times$ less power compared to other reported implementations in the literature.
    \item The sparse-aware ADC provides $>$20$\times$ reduction in ADC conversion energy by converting only significant JPEG compressed samples, which typically constitute $\le5\%$ of the total captured analog samples.
    \item We proposed an \textit{Automatic calibration} scheme to update the Q-matrix during decoding, which negates the process variations and effect of parasitics in switched-capacitor design
    \item Furthermore, the power expended for communication/transmission is lowered by compressing the signal in the analog domain. With the reduction in the number of bits, the communication energy reduces drastically.
 \end{itemize}
 
 \subsection{Organization of paper}
The remainder of the paper is organized as follows. Section II discusses, the theoretical and mathematical background of JPEG compression. In Section III, the different design considerations are studied, including the comparison between Voltage mode and Current mode implementation, capacitor selection for the DCT matrix and Quantization matrix(Q-Matrix), metrics of image performance evaluation-PSNR, variation tolerance using Q-sense and lastly the overall Input image processing done for testing purpose. Section IV presents the proposed architecture for the dAJC IC, illustrating the circuit-level implementation of the various JPEG compression blocks such as 2D-DCT, quantization, zig-zag traversal, run-length encoder, and sparsity-aware ADC, along with the overall non-overlapping phase generation circuit to perform the JPEG compression. Section V provides the overall measurement results for the dAJC IC.  Section VI compares the proposed design with the related works reported in the literature. Section VII, concludes the paper.

\section{JPEG Compression}
\begin{figure}[htp]
   \centering
    \includegraphics[width=0.48\textwidth]{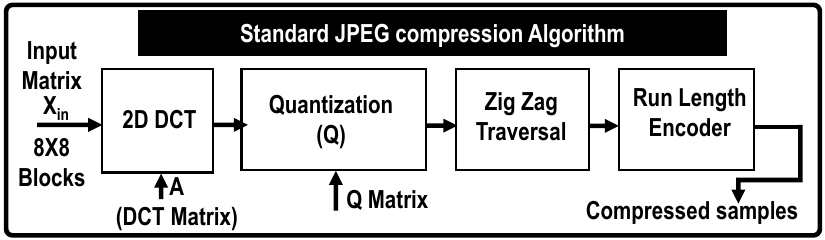}
    \caption{Standard JPEG compression algorithm and its constituting blocks.}
    \label{fig:Simple_JPEG}
    \vspace*{-0.2in}
\end{figure}

This section discusses the basic overview of the operations constituting the Joint Photographics Experts Group (JPEG) compression \cite{125072}method, highlighting underlying operations such as Discrete Cosine Transformation(DCT), Quantization, Zig-zag traversal, and Run length encoding. Fig. \ref{fig:Simple_JPEG} shows the overall JPEG compression diagram with various blocks highlighted.

\subsection{Discrete Cosine Transformation(DCT)} 

The first step of performing JPEG compression is DCT\cite{khayam2003discrete},\cite{lam2000mathematical}. DCT performs the transformation of the image data from the time domain to the frequency domain. 
\begin{figure}[htp]
   \centering
    \includegraphics[width=0.48\textwidth]{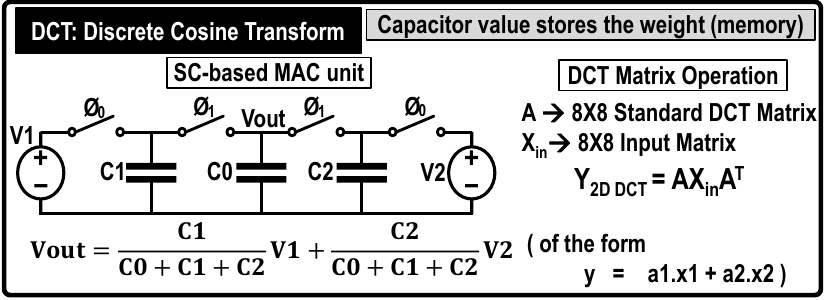}
    \caption{DCT operation consists of matrix multiplication requiring MAC operations. The schematic shows the MAC operation being implemented by a Switched-capacitor-based circuit.}
    \label{fig:DCT_basic}
\end{figure}
Because of the sparsity of the DCT basis, in the frequency domain, the image information is sparse, that is image information in the frequency domain consists of more 0's compared to the number of 1's, for a particular frequency. Hence the original image data can be represented with fewer samples\cite{saha2000image}. Mathematically, DCT is the matrix multiplication of the standard DCT matrix with the input matrix(input data).

In Fig. \ref{fig:DCT_basic}, 'A' represents a standard 8$\times$8 DCT matrix and $X_{in}$ represents the 8$\times$8 input matrix. In practice, an 8$\times$8 matrix is selected, although selecting a larger size matrix of N$\times$N where $N>9$, gives better compression but it takes a great deal of time to perform DCT calculations, creating an unreasonable trade-off between the overall architecture complexity and the speed of computation. The DCT operation is given by Eq. \ref{DCT_eq}: 
\begin{equation} \label{DCT_eq}
[Y]_{2D-DCT} = [A].[X_{in}].[A^T]
\end{equation}

From the equation, we can infer that the 2D-DCT operation is the result of matrix multiplication. Hence Multiplication and Accumulation(MAC) are the two operations being performed to generate the Y-matrix. The MAC operation realization using a Switched-Capacitor circuit is shown in Fig. \ref{fig:DCT_basic} as an SC-based MAC unit. In phase $\phi_0$, the voltage V1 and V2 are sampled on the capacitors C1 and C2 respectively. In the next phase $\phi_1$, the voltages on capacitors C1 and C2 are accumulated onto the capacitor C0(accumulation capacitor), generating the $V_{out}$ voltage which is the resulting MAC operation of voltages V1 and V2. The multiplication coefficient a1 and a2 can be controlled by adjusting the C0, C1, and C2 values. Hence the MAC operation is realized using the switched-capacitor circuit. Similarly, performing the 2D DCT operation needs multiple such MAC units which can be realized using SC-MAC units that operate synchronously in a parallel architecture.

\subsection{Quantization}
Quantization performs element-by-element division of the output of the 2D-DCT block by the Q-matrix \cite{watson1993dct}. Depending on the amount of compression and the quality of the image required, the standard Q-matrix varies. 

\begin{figure}[htp]
   \centering
    \includegraphics[width=0.48\textwidth]{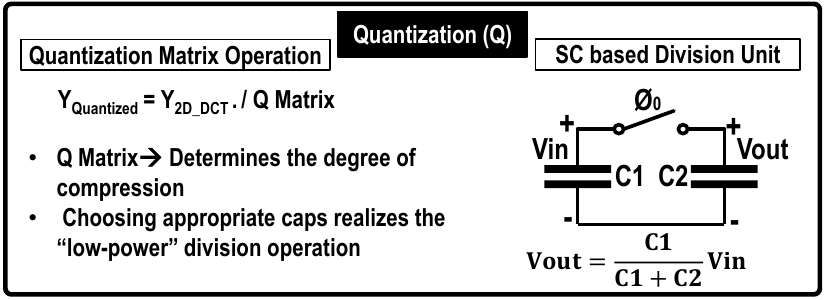}
    \caption{Quantization operation involving element-wise division operation. Division operation  implementation using switched capacitor circuit}
    \label{fig:Basic_Q}
    \vspace*{0in}
\end{figure}

For example, $Q_0$ provides maximum compression and minimum image quality. However, $Q_{90}$ provides the minimum compression and better image quality. To achieve a trade-off between the image quality and the degree of compression we have implemented $Q_{50}$ matrix. The output of the Q-matrix is an element-wise division operation given by the following equation\ref{DCT_operation_referred}:
\begin{equation} \label{DCT_operation_referred}
[Y]_{Quantized} = \frac{[Y]_{2D DCT}}{Q_{matrix}}
\end{equation}

The division operation is realized using a switched-capacitor circuit as shown in Fig. \ref{fig:Basic_Q}. To perform division, an output of the 2D DCT matrix is first stored in the capacitor C1(represented as Vin). In the next phase, the charge of capacitor C1 is dumped to capacitor C2 such that the resulting voltage(represented as Vout) is the required division operation such that the divisor value is determined by the capacitor C1 and C2 values.
\subsection{Zig-Zag traversal}

Zig-Zag traversal receives the 8$\times$8 matrix from the quantization block and serializes the matrix such that the most significant samples appear in the beginning followed by a long chain of insignificant samples(which can be ignored without degrading the image quality).

\begin{figure}[h]
   \centering
    \includegraphics[width=0.48\textwidth]{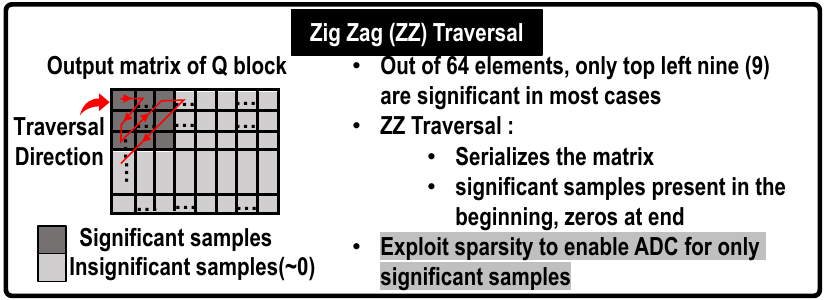}
    \caption{Zig-Zag block showing the arrangement of significant and non-significant samples.}
    \label{fig:Simple_zig_zag}
    \vspace*{0.0in}
\end{figure}
Out of the 64 samples in the quantized matrix, in most cases, only the top left samples are the most significant samples and carry the most information about that particular frame as shown in Fig. \ref{fig:Simple_zig_zag}. However, this block does not perform the final compression since the output of the block still consists of 64 samples.

\subsection{Run Length Encoder(RLE)}
The Run length encoder block receives the serialized output from the zig-zag traversal block such that the first few samples contain significant information about the frame followed by low amplitude insignificant samples. It performs the actual compression by generating an output containing significant samples followed by the count of the number of insignificant samples.
\begin{figure}[h]
   \centering
    \includegraphics[width=0.47\textwidth]{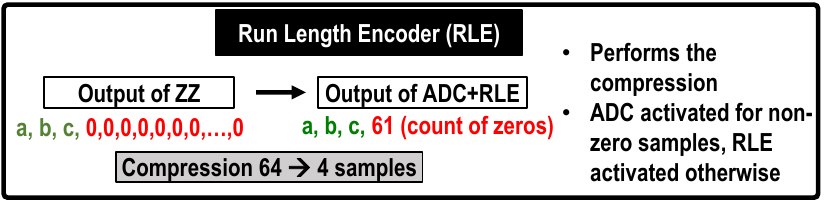}
    \caption{Simplified operation of the Run Length Encoder block performing compression }
    \label{fig:simple_RLE}
    \vspace*{-0.1in}
\end{figure}

As shown in Fig. \ref{fig:simple_RLE}, the output of the RLE contains the significant samples a,b, and c followed by 61 insignificant samples. Hence performing a 16$\times$ compression. Depending on the samples from the zig-zag traversal block the sparsity-aware ADC is activated only for the significant samples while the rest of the time the RLE counts the number of insignificant samples, thereby also reducing the ADC power consumption by exploiting the sparse nature of the input from the zig-zag traversal block.

\section{Design Considerations}
\textcolor{black}{This section provides valuable insights into the different architectural explorations performed to realize an optimized design (taking into process variations, minimum area), the performance metrics used for evaluation, the choice of capacitor for the 2D-DCT matrix and quantization matrix}, voltage mode vs current mode architecture and the process of generating the input image for testing purpose.

\subsection{Voltage Mode versus Current mode Comparison} 
The voltage mode implementation of the DCT calculation is implemented using capacitors, voltage sources, and switches. It is based on the principle of charge redistribution across the capacitors. Hence major power consumption in the circuit is due to the losses across the real switch, clocking circuitry, and the losses while charging and discharging the capacitors. However, the current mode of implementation will require multiple current sources to discharge/charge the voltage on a capacitor to multiply/divide the input with the required value of DCT coefficients. It is based on the principle of integrating the voltage over a capacitor using a current source. So to implement various DCT coefficients and perform element-wise division, a large number of accurate current sources are required leading to an increase in the total power. Although accurate current copying techniques can be employed, they reduce the overall dynamic range of the intermediate signals, making the overall processing more prone to noise. Considering all of these, the voltage mode implementation is chosen as it requires lower power to perform the DCT implementation compared to its current mode implementation. The trade-off for the voltage mode of implementation is the area overhead of implementing the coefficients using capacitors and the effect of the parasitic capacitor on the actual capacitors(which is tackled using Q-sense and calibration mechanisms to be discussed later).  
\begin{figure}[h]
   \centering
    \includegraphics[width=0.47\textwidth]{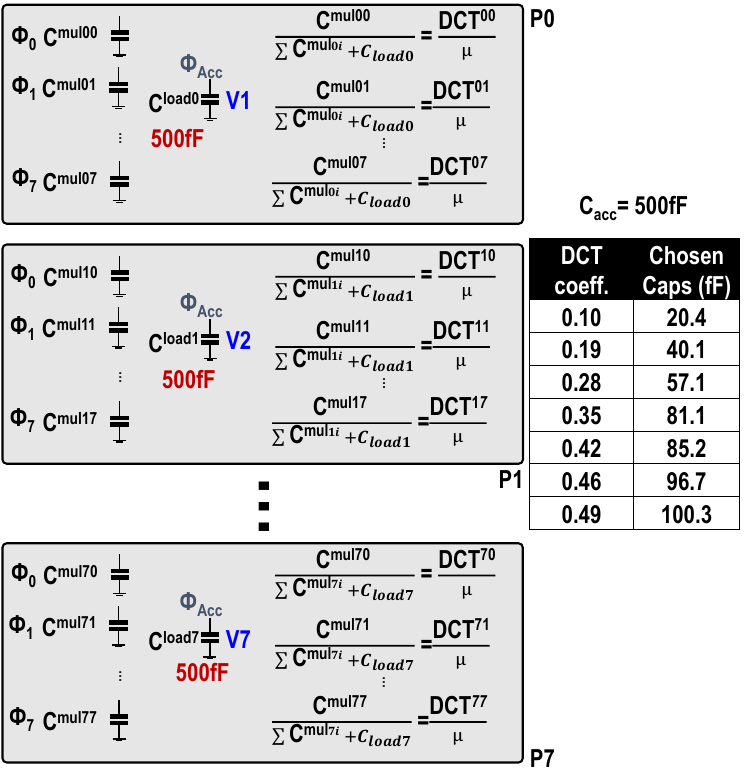}
    \caption{Design Considerations for Selecting the Capacitor Values to Implement 2D-DCT Coefficients}
    \label{fig: DCT_cap_choice}
    \vspace*{0in}
\end{figure}

\begin{figure}[h]
   \centering
    \includegraphics[width=0.47\textwidth]{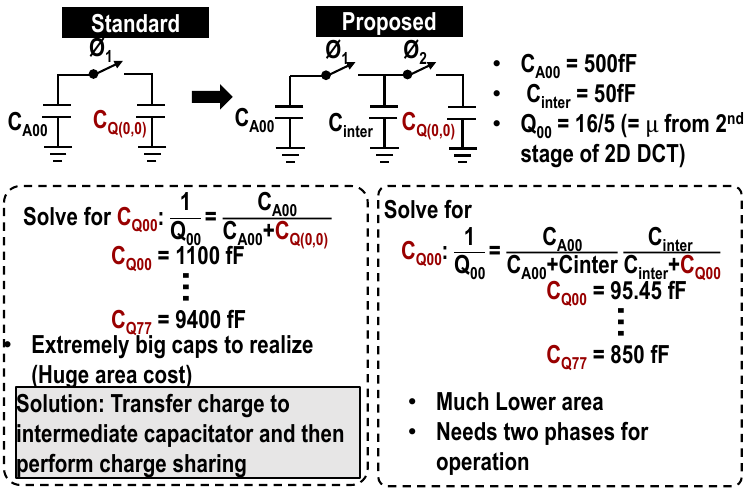}
    \caption{Design Considerations for selecting the Quantization Capacitor values}
    \label{fig:quant_cap}
    \vspace*{-0in}
\end{figure}

\begin{figure*}[htp]
   \centering
    \includegraphics[width=0.91\textwidth]{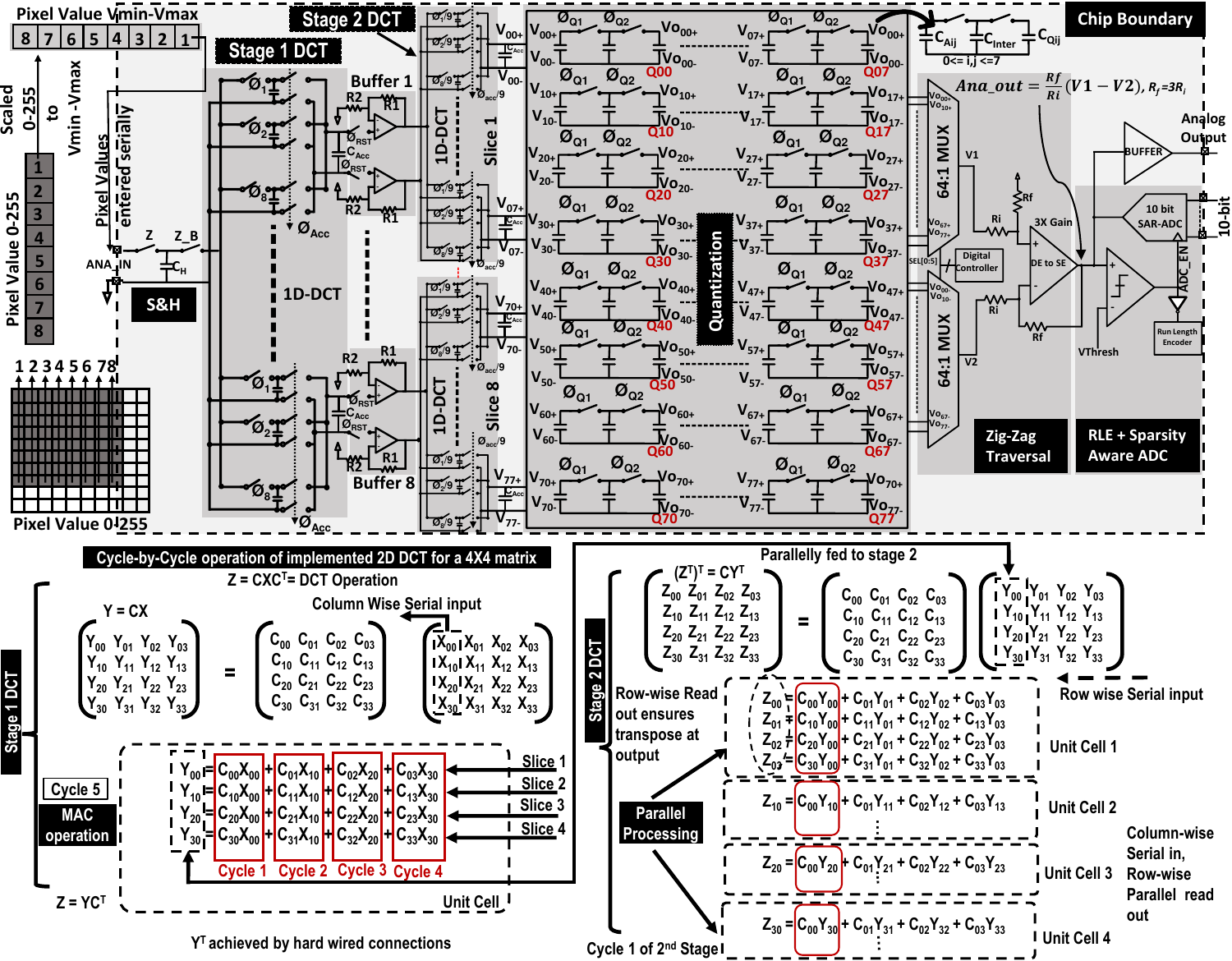}
    \caption{The overall circuit level diagram of the 2D-DCT implementation. (Bottom Figure) Demonstrates the Cycle-by-Cycle operation of the implemented 2D DCT for a 4$\times$4 Matrix. }
    \label{fig:2d_dct_ckt}
    \vspace*{-0.2in}
\end{figure*}

\subsection{Metrics for Image Performance evaluation-PSNR and SSIM}
To evaluate the quality of each frame we use standard performance evaluation metrics: Peak Signal Noise Ratio(PSNR) and SSIM(Structural Similarity Index)\cite{5596999}. PSNR helps evaluate the overall quality of the frame and the higher its value the better the recovered frame quality. SSIM is the more approximate measure of image quality but is a better fit for the human perception of the image. An image can be easily perceived by the human eye if the frame has an SSIM $>$ 0.85. In the design, the image quality can be controlled by controlling the intermediate signal levels.

\subsection{Choice of Capacitor for DCT Matrix}
The capacitor value is chosen based on the DCT coefficient values. The value of the capacitor acts as a memory as well as helps in the realization of the coefficient values. The value of the capacitor is determined by the DCT coefficients needed to be implemented as well as the overall attenuation introduced during computation of the MAC operations and moving charge from one capacitor to another. In Fig. \ref{fig: DCT_cap_choice}, considering a value of 500fF as the accumulation capacitor, various DCT coefficients are shown, such that to implement the minimum DCT coefficient value of 0.1, the capacitor value is chosen to be 20.4fF and for the DCT coefficient of 0.49, the value of the capacitor is 100fF. From the calculation, the minimum capacitor value is chosen such that to implement the smallest DCT coefficient the capacitor is as small as possible to reduce the capacitor size for a higher value of the DCT coefficient and reduce the overall area, simultaneously taking precautions such that the minimum capacitor is not affected by the parasitic capacitors. Also, the selection of an accumulation capacitor is such that the overall divided value of the minimum voltage is above the noise floor of the system. Similarly, for each 8$\times$1 block, several such sampling capacitors $Cmul_{ij}$ where i,j varies from 0-7 are implemented.

\subsection{Choice of Capacitor for Quantization Matrix}
A quantization matrix is used to divide the output of the 2D-DCT block element-wise using a Q-matrix to provide a certain degree of compression.

 $Q_{00}$ corresponds to the highest degree of compression whilst ruining the quality of the image. Similarly, $Q_{100}$ corresponds to the lowest compression but highest quality image. In this implementation, $Q_{50}$ has been realized. The division is performed using a switched-capacitor-based circuit, as explained previously. The purpose of the capacitor value is similar to that of the DCT operation where the coefficient or the division value is stored in the capacitor value. In the first implementation, the division operation is realized using only one stage of division operation as shown in Fig. \ref{fig:quant_cap}. However, this implementation requires large capacitors consuming a larger area. However, in the final on-chip implementation the division is achieved by a two-stage division operation. To perform the division, in the first stage, the charge is transferred to an intermediate capacitor and then transferred to another cap $C_{Q00}$ to perform the overall division operation. Hence by this implementation, the overall area can be reduced significantly by 10$\times$ for each division operation in the quantization matrix providing a significant reduction in area. However, the final implementation requires a total of two phases to perform the overall elementwise division operation.

\subsection{Input Image processing}
The targeted application works on the principle that the sensed image pixels are represented in terms of voltage, otherwise known as the analog camera. The camera OV7670 is an analog camera module by Omnivsion, which generates the image in the form of an analog signal, preferably voltage. First, the image is sensed by the image sensor in the form of M$\times$N frame size. Then the internal hardware serializes the M$\times$N matrix either column-wise or row-wise in blocks of M$\times$1 or N$\times$1 in the set of 8$\times$8 matrix, where each element of the matrix is a voltage level representing a particular pixel value. To emulate the functionality of an analog camera, a grayscale image is taken as an input. 

Then the grayscale image values are mapped onto a certain voltage scale of Vmin to Vmax, such that a pixel value of '0' maps to a voltage of Vmin and a pixel value of '255' maps to a pixel value of Vmax, using MATLAB. Then the pixel values are serialized row-wise or column-wise and sent to the test IC using an arbitrary waveform generator(AWG) as a continuous-time signal of a certain data rate. The overall architecture is serial hence the input is taken serially to save on the number of input ports and avoid synchronization between the various IO ports.

\section{Proposed Architecture for d-AJC}
The overall architecture is chosen based on the previous discussions. A voltage mode implementation along with a two-step division in the quantization matrix while keeping the overall implementation switched-capacitor-based has been implemented on the IC.
As shown in Fig.\ref{fig:2d_dct_ckt}, the overall architecture consists of the 2D-DCT processing block using SC-MAC, quantization matrix, zig-zag traversal block, and RLE + sparsity-aware ADC. In the overall processing, there are losses due to the charge sharing while moving charges from one stage to another. To compensate for the losses and keep the signal on an acceptable level at the end of processing, voltage buffers are introduced in between 1D-DCT and 2D-DCT blocks and after the quantization operation. The topology selection and the value of buffer gains are also discussed in this section. 

\subsection{2D-DCT}
The 2D-DCT operation comprises two-stage matrix manipulations as shown in the bottom of Fig.\ref{fig:2d_dct_ckt}. The output of the 2D-DCT block (Z) can be expressed as, Eq. \ref{DCT_operation}  
\begin{equation} \label{DCT_operation}
[Z]_{2D-DCT} = [C].[X_{in}].[C^T]
\end{equation}
where that C represents the DCT matrix, $X_{in}$ represents the input matrix and $C^T$ represents the transpose of the DCT matrix. 
The \textit{Stage1 DCT} block takes column-wise serial input of the image and performs MAC operation with the DCT matrix. 
The intermediate result $[C] \times [X_{in}]$ is obtained at the output of the \textit{Stage1 DCT} block.
Fig.\ref{fig:2d_dct_ckt} (bottom) shows the cycle-by-cycle 2D DCT operation for a  $4 \times 4$  matrix. For a $4 \times 4$ 2D DCT computations, 4 slices with capacitors are required to store the required coefficient values.
The different partial products are calculated and stored on the corresponding capacitors in slices in subsequent non-overlapping cycles. 
In the fifth cycle, the partial products are summed up and stored on a larger accumulator capacitance. 
Note that the reset phase of the accumulator capacitor is merged with one of the sampling phases as they are fundamentally non-overlapping.
The outputs of \textit{Stage1 DCT} are fed parallelly to \textit{Stage2 DCT} which performs residual a $4 \times 4$ matrix multiplication of the result ($[C] \times [X_{in}]$) with the transpose of the DCT matrix ($[C^T]$). 
Each partial product computation in \textit{Stage2 DCT} needs to wait for the output of \textit{Stage1 DCT}, and thus we require a total of $(4 \times 4) + 4$ cycles to get the output of the  2D DCT block. 
The outputs of the 2D DCT block are read row-wise to emulate the transpose of the DCT matrix.

\begin{figure}[htp]
   \centering
    \includegraphics[width=0.4\textwidth]{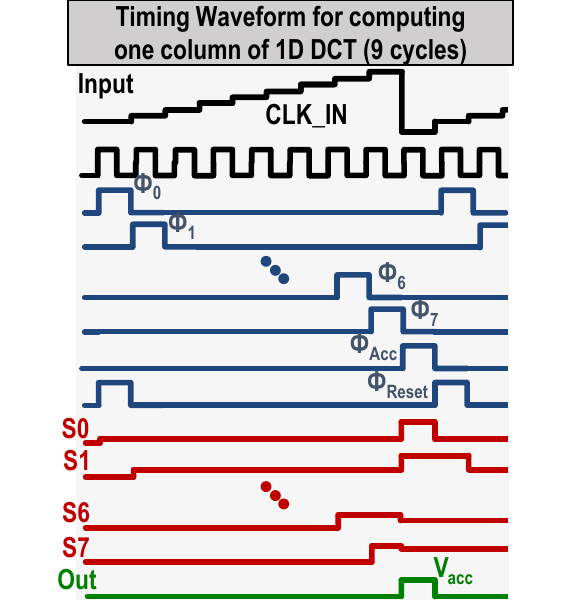}
    \caption{Timing Waveform for 1D-DCT operation}
    \label{fig:1D_clocking}
    \vspace*{0in}
\end{figure}

\begin{figure}[htp]
   \centering
    \includegraphics[width=0.35\textwidth]{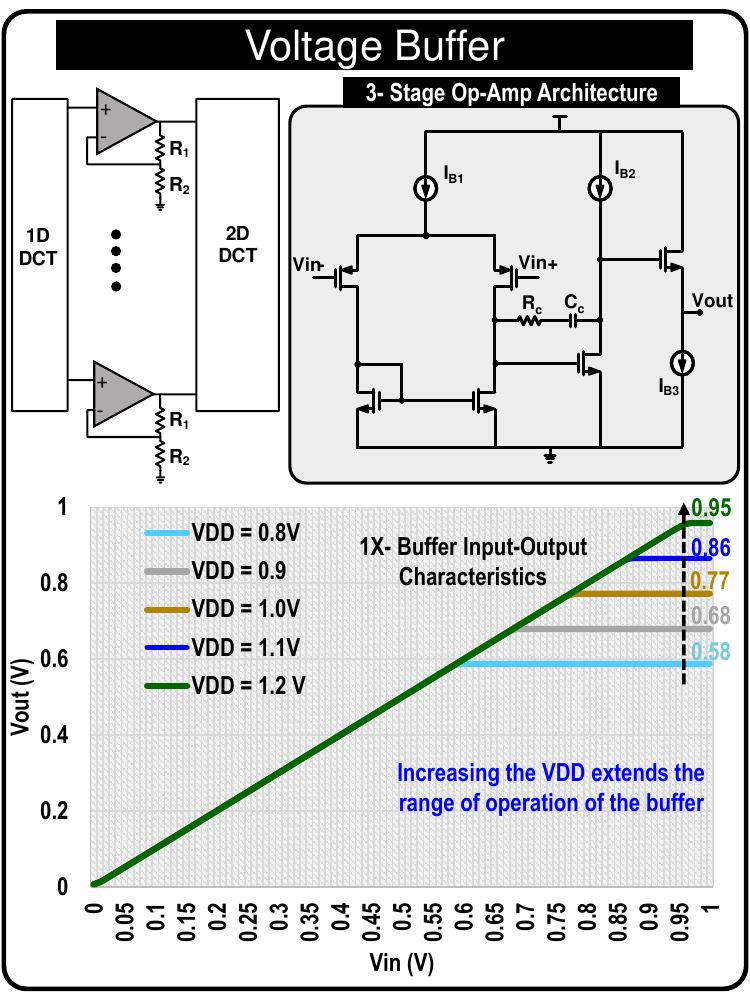}
    \caption{Op-amp circuit diagram used for 2$\times$ gain amplifier. Plots showing the output of the op-amp in unity gain configuration.}
    \label{fig:inter_buf_ckt }
    \vspace*{0.1in}
\end{figure}

Similarly, for the JPEG algorithm with $8 \times 8$ Matrix, inputs are fed in $8 \times 1$ form such that each set of partial product calculations takes 1 cycle and thereby requires 8 cycles to perform $8 \times 8$ multiplication, and one cycle is required to accumulate the voltage on the accumulator capacitor. The accumulator capacitor is then reset in one of the sampling phases before feeding to the following stage. 
The output of the \textit{Stage1 DCT} is fed to the \textit{Stage2 DCT} through an intermediate buffer that provides signal gain and helps to improve the SNR of the system, discussed in detail later.
Consequently, 1st stage of DCT takes a total of nine data cycles, 8 cycles for sampling, and 1 for accumulation to produce the output for each input column as shown in Fig.\ref{fig:1D_clocking}. 2nd stage DCT block runs on a slower clock than the 1st stage DCT block to sample the output of the 1st stage DCT. Hence the total time to perform 2D DCT of $8 \times 8$ block is 72 cycles (9 cycles $\times$ 8). 
The operations such as sampling, accumulation, and resetting are done using a non-overlapping phase generator circuit which generates the various phases depending on its input clock. Because of implementation using switched and various clock phases, there might be clock coupling issues which are taken care of by using transmission gates as the switch and doing the processing differentially to cancel out common mode switching noise.

\subsection{Intermediate buffer}
The overall architecture is based on the principle of charge sharing. So, while sending the output of the 1st DCT stage to the input of 2nd DCT stage,
the charge has to be transferred from one small capacitor at the DCT output to the input of the 2D DCT block which has a higher input capacitor which leads to the addition of the attenuation. This attenuation is introduced due to charge sharing is an intentional attenuation introduced due to charge sharing which is utilized to avoid saturation in the following processing blocks. While charging the 2D DCT input from the 1D-DCT there is an attenuation of 0.5, which is introduced by charge sharing, and hence the requirement of a voltage buffer with a 2$\times$ gain rises. Initially, to implement the buffer a passive voltage multiplier using a series-parallel combination of capacitors was used. But it suffers from the problem of leakage, the size of the capacitor increases the overall area and lastly stacking more than one capacitor does not provide a proportional multiplication factor. Various other active voltage buffers\cite{carvajal2005flipped} were tested such as the Flipped Voltage follower, folded flipped voltage follower, and common drain however linearity was a concerning issue considering that the output of the buffer has to be precisely 2$\times$ without any distortion in the value achieved. To get the linearity, an op-amp-based buffer has been used in the current implementation. The buffer consists of three stages. The first two stages provide the required gain and the last stage is a voltage follower to drive the high input capacitance of the 2D DCT Stage. The overall op-amp has been selected to be a PMOS input stage based since the DCT stages produce an output having a lower common mode voltage as a result of arithmetic calculations; it also cuts down the power since the operating range is reduced. Fig.\ref{fig:inter_buf_ckt }, shows the overall placement of the buffer between 1D-DCT and 2D-DCT and the PMOS-based op-amp circuitry. The supply range can be increased to increase the input voltage range. The gain of the buffer can be controlled by controlling the R1 and R2 resistors, which helps cancel out the process variations in the resistors implemented.

\subsection{Quantization Block}
The quantization block performs the element-wise division of the 2D-DCT matrix with the quantization matrix using a switched capacitor-based division operation.
\begin{figure}[htp]
   \centering
    \includegraphics[width=0.47\textwidth]{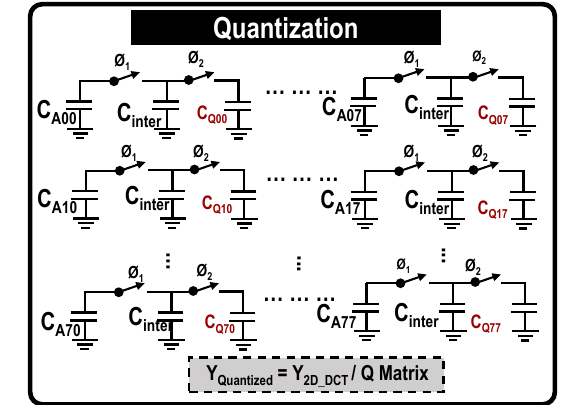}
    \caption{Circuit Level diagram of the Quantization block showing 2-step division operation}
    \label{fig:quant_ckt}
    \vspace*{-0in}
\end{figure}
The charge redistribution among capacitors $C_{Aij}$, $C_{inter}$, and $C_{Qij}$ computes the quantization, which is a division operation. The ratio of capacitors $C_{Aij}$, $C_{inter}$, and $C_{Qij}$ stores the values of Q-matrix operating as a memory, where, 0 $\le$ i $\le$ 7, 0$\le$  j $\le$ 7. As discussed before the division operation in the quantization block is performed using a two-step division process to reduce the capacitor size required. However, that leads to the requirement of two phases to perform the division operation. Also, care has been such that $C_{Qij}$ is 90fF to reduce the effect of parasitics and have the ability to drive the capacitance of the following 64:1 analog MUX.

\subsection{Zig-Zag Traversal Block}
The output of the Quantization block consists of significant samples and non-significant samples in a matrix form. The Zig-zag traversal block arranges the elements of the matrix in a serial manner such that significant samples are at the start of the series and then followed by lower-valued samples. 
\begin{figure}[htp]
   \centering
    \includegraphics[width=0.47\textwidth]{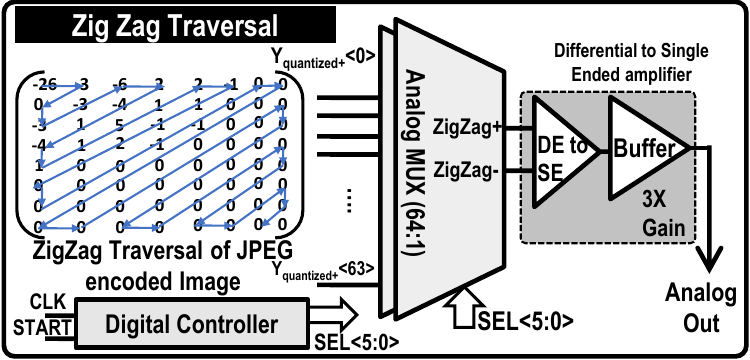}
    \caption{Circuit Level diagram for the Zig-Zag Traversal Block}
    \label{fig:zig_zag_ckt}
    \vspace*{0in}
\end{figure}
The traversal as shown in Fig.\ref{fig:zig_zag_ckt} is implemented using a digital controller and 64:1 analog MUX. The output of the zig-zag traversal clock is a serial differential signal. Since the previous operations were differential, the differential signal is converted into a single-ended output to send it to the single-ended sampler for a decision of significant and non-significant samples. A 3$\times$ gain buffer is added to compensate for intentional attenuation added in the chain. This buffer is also op-amp based whose gain is controlled using the feedback resistors.

\subsection{Run length Encoder Block with ADC}
After serialization by the zig-zag traversal block, the actual compression has not yet been performed.
\begin{figure}[htp]
   \centering
    \includegraphics[width=0.47\textwidth]{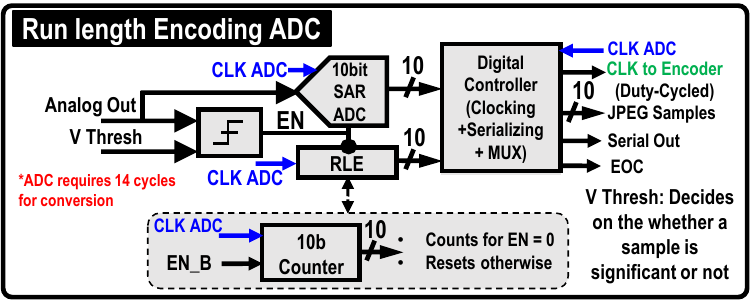}
    \caption{Circuit level diagram for the Run-Length Encoder along with the Sparsity-Aware ADC}
    \label{fig:rle_ckt}
    \vspace*{-0in}
\end{figure}
As shown in Fig.\ref{fig:rle_ckt}, the run-length encoder block \cite{van1976extended}takes in the analog input from the zig-zag traversal block, and depending on the amplitude of the sample, generates outputs containing the significant samples followed by the count of the samples. The functionality of the RLE is realized using a sampler that samples the output coming from the zig-zag traversal block and compares it with a threshold( indirectly controls the image quality). If the sample is significant then the sample is digitized at the ADC otherwise the RLE is activated to count the number of insignificant samples. The sampler implemented is a PMOS-based sampler selected based on the voltage level available at that stage and its reference voltage is controlled to control the degree of compression.
\begin{figure}[htp]
   \centering
    \includegraphics[width=0.47\textwidth]{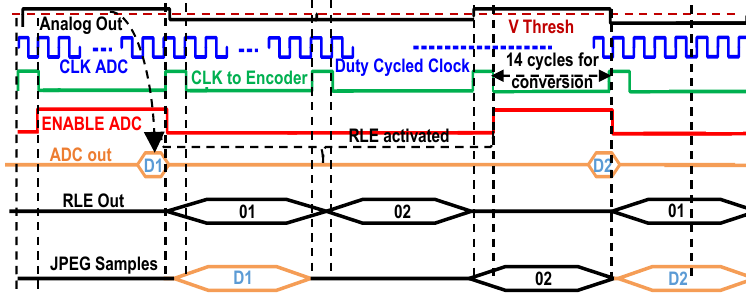}
    \caption{Timing diagram showing the overall operation of the sparsity-aware ADC}
    \label{ref:timing_control_adc}
    \vspace*{-0.2in}
\end{figure}

\subsection{Operation of the Sparsity-Aware ADC}
The timing control for the ADC is illustrated in Fig.\ref{ref:timing_control_adc}. Initially, when the incoming analog sample exceeds the VThresh voltage threshold, indicated as 'Analog out', the ADC is activated. Operating at a higher speed than the sampling clock, the ADC produces the digitized output D1 after \textbf{14} cycles.

In cases where analog samples fall below the VThresh threshold, the ADC is deactivated while the RLE block is activated. This block executes run length encoding and accumulates the count of insignificant samples until the subsequent significant analog sample is detected. Upon arrival of the next significant sample, the RLE outputs the count of insignificant samples, such as '02' in Fig. \ref{ref:timing_control_adc}. Overall, the block's output comprises the digitized JPEG samples, followed by the count of insignificant samples.

\begin{figure}[htp]
   \centering
    \includegraphics[width=0.40\textwidth]{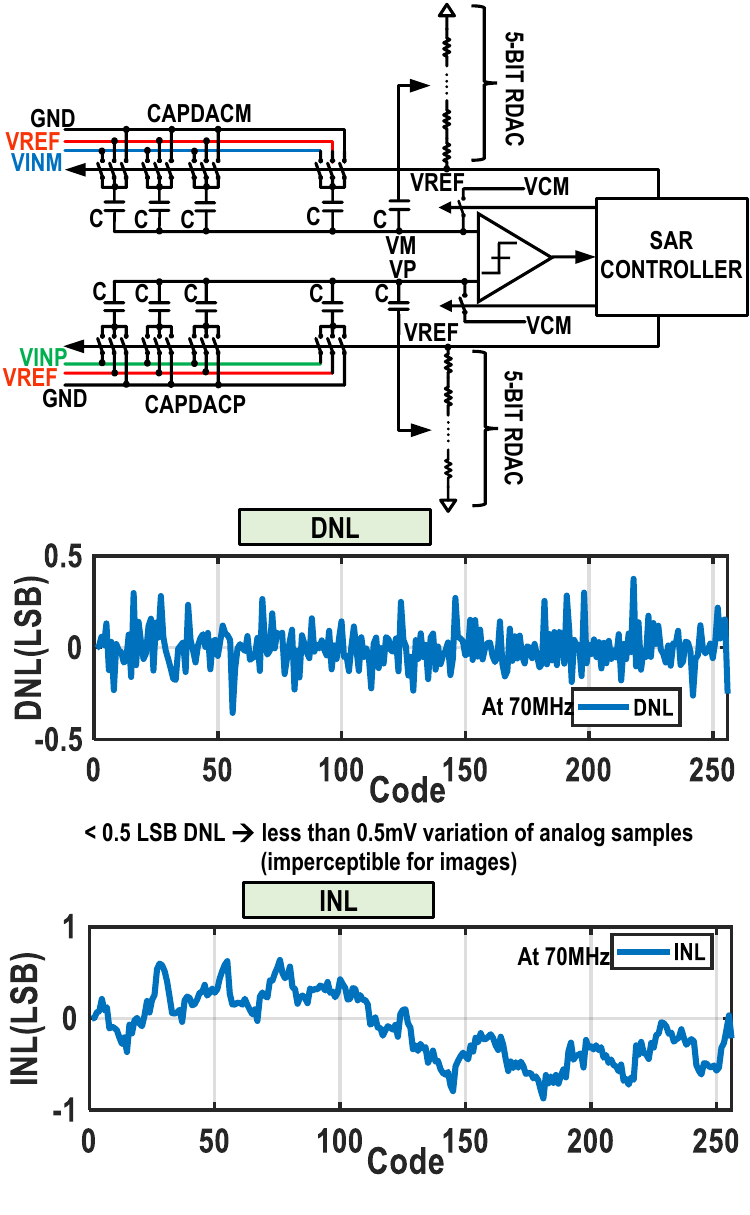}
    \caption{Circuit level diagram of the 10-bit SAR ADC along with the INL and DNL plots.}
    \label{fig:adc_ckt}
    \vspace*{-0.22in}
\end{figure}

\subsection{10-bit SAR ADC}
A 10-bit Successive Approximation based differential ADC is implemented to convert the significant analog samples for digitization and further processing. The block diagram of the SAR-ADC is shown in Fig. \ref{fig:adc_ckt}. A 5-bit charge scaling capacitive DAC ( MSB 5 bits) and a 5-bit voltage scaling resistive DAC ( LSB 5 bits) are used to implement a 10-bit hybrid DAC. Since the capacitor matching is better than that of the resistors, the DAC has better accuracy in the MSBs. Also, the LSBs are monotonic. The unit capacitor value in the design is 9.8$f$F. A thermometric decoder is used to select the capacitor array to have a lower DNL. The designed ADC can be configured as a single-ended ADC by connecting the VREF and VINM of CAPDACM to the ground. The ADC is designed following charge redistribution \cite{mccreary_all-mos_1975} on the top plate of the capacitor array. The first 3 clock cycles are used to sample the signals. During the sample and tracking phase, the top plate of the capacitors is connected to the common mode voltage, VCM, and the bottom plate is connected to the signal. The charge stored in the capacitor array is proportional to the signal in the sampling and tracking phase. The next 10 clock cycles are used for conversion. The conversion starts by connecting the MSB capacitor to the VREF.

\begin{figure*}[htp]
   \centering
    \includegraphics[width=1\textwidth]{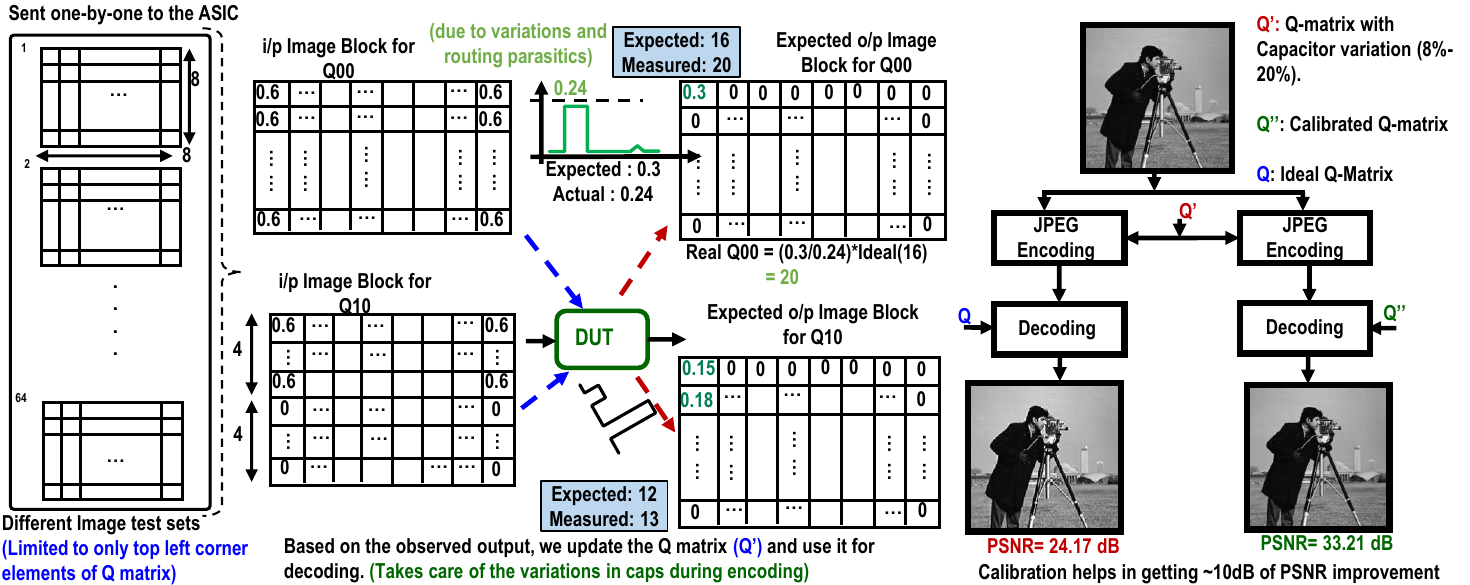}
    \caption{Overall calibration mechanism of the inverse-Q Matrix to counter process variation and effect of parasitics in the IC.  }
    \label{fig:q_calibration}
    \vspace{-0.1in}
\end{figure*}

\subsection{Non-Overlapping clock generator}
\begin{figure}[htp]
   \centering
    \includegraphics[width=0.4\textwidth]{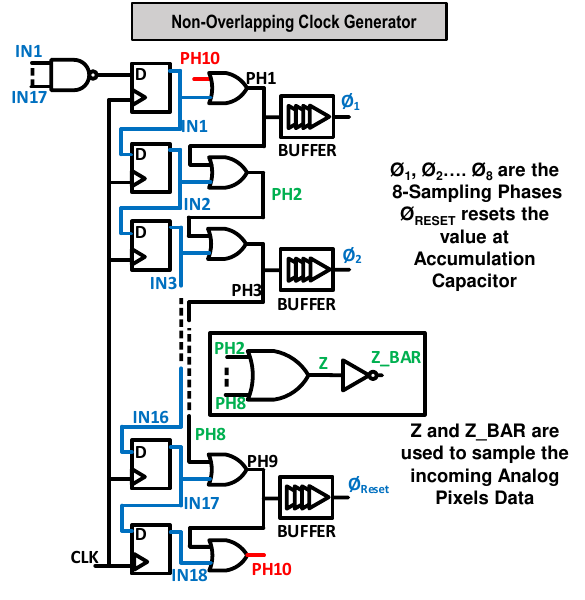}
    \caption{Circuit level diagram of the Non-Overlapping Clock generator, showing the clock phases generated to perform the 2D-DCT operation}
    \label{fig:non_ov_ckt}
    \vspace*{-0.3in}
\end{figure}

The 2D DCT block executes sampling, accumulation, and division operations, each requiring distinct non-overlapping clock phases\cite{hari2011low}. 
These phases are generated by a non-overlapping clock generator block, as depicted in Fig. \ref{fig:non_ov_ckt}.
This block takes an input clock and divides it to produce a slower clock, used to generate non-overlapping phases. The duration of non-overlap at rising and falling edges is controlled by delay in the feedback path.
To generate the 9 phases necessary for DCT operations, the clock generator initially requires 9 cycles. Subsequently, these phases are utilized by the DCT block for processing. The timing of these clock phases, as explained in Fig. \ref{fig:1D_clocking}, ensures proper synchronization and execution of the DCT operations.

\begin{figure}[htp]
   \centering
    \includegraphics[width=0.31\textwidth]{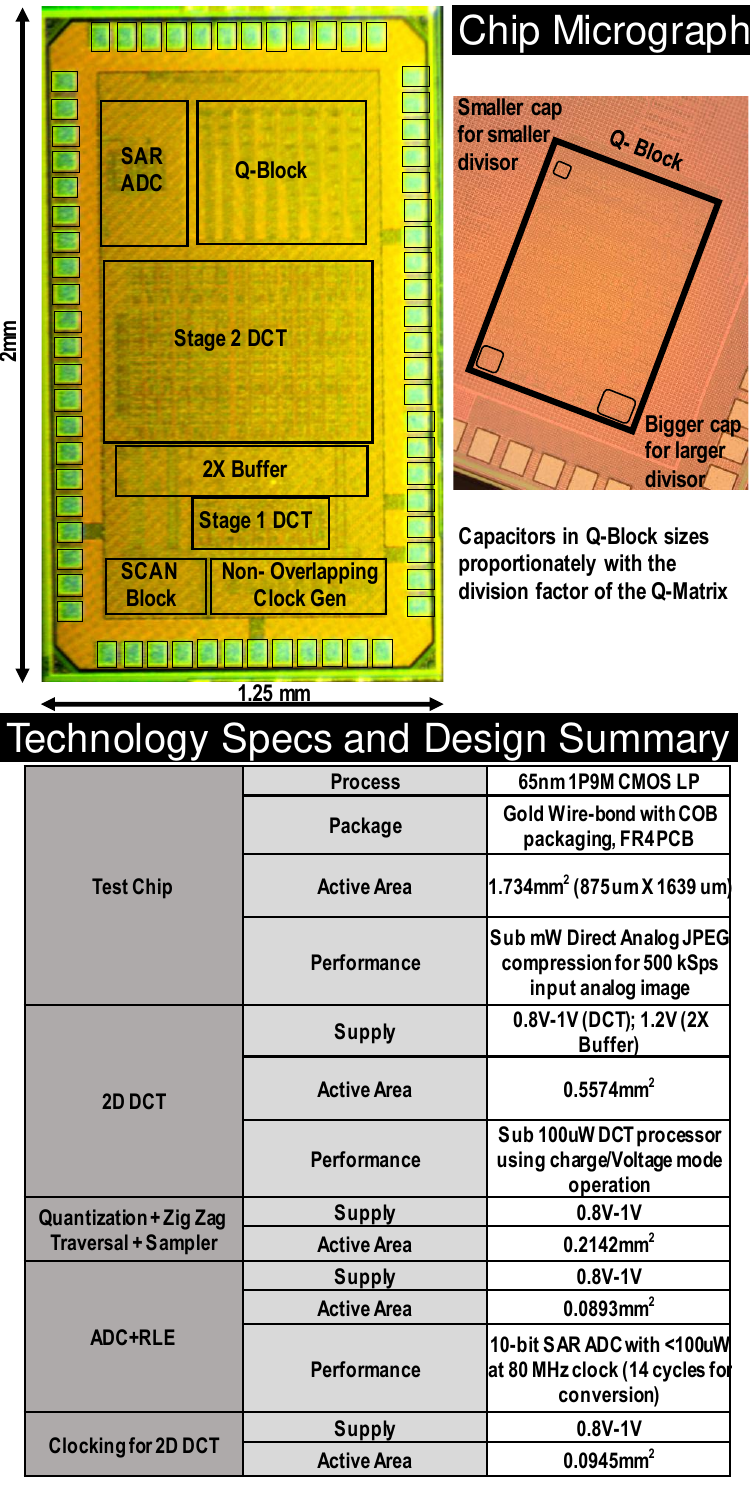}
    \caption{Die Micrograph and Highlighted region with the Q-matrix implemented On-Chip}
    \label{ChipDie}
    \vspace*{0in}
\end{figure}

\begin{figure*}[htp!]
   \centering
    \includegraphics[width=0.84\textwidth]{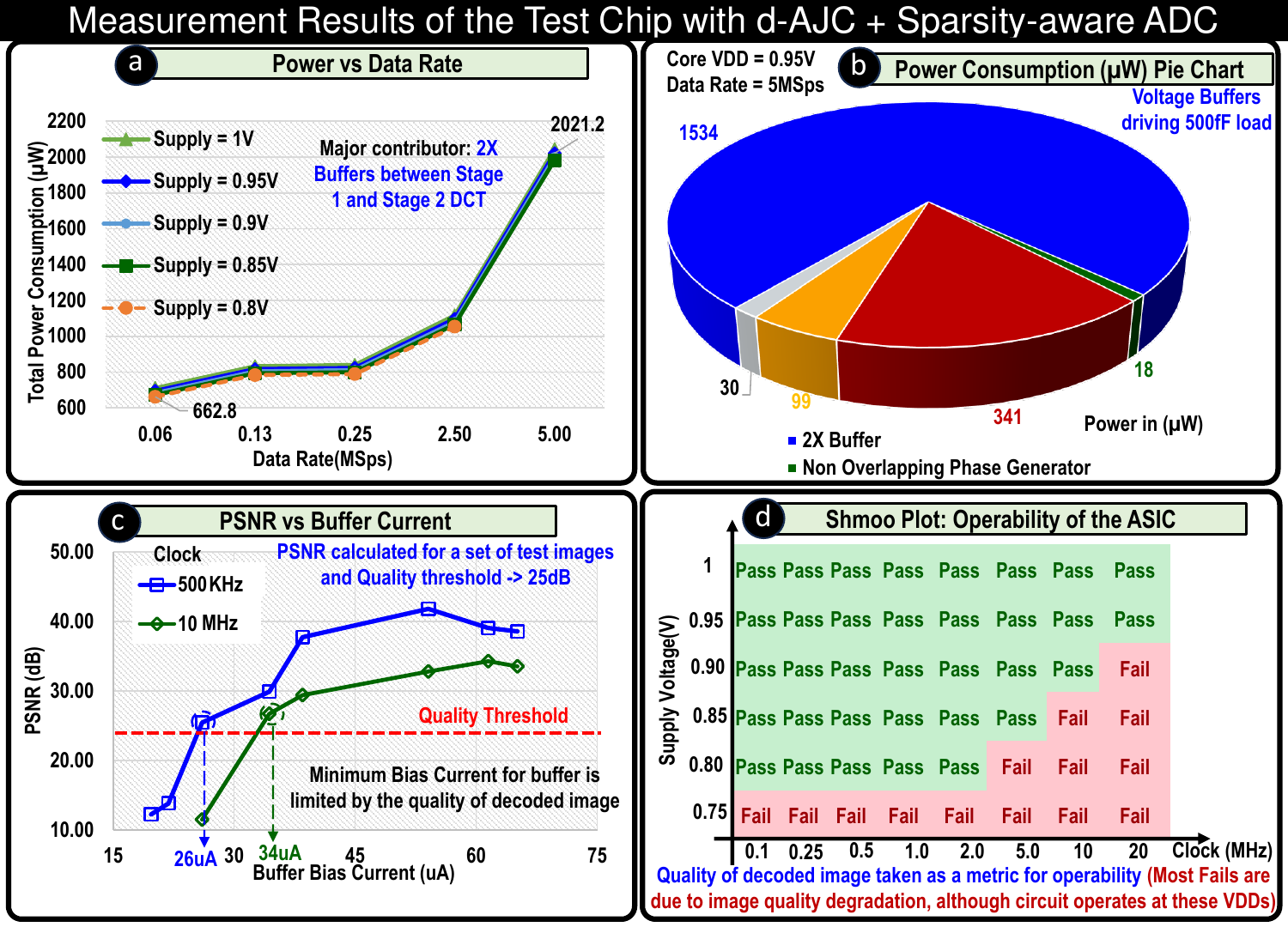}
    \caption{Overall Measurement Results (a) Power vs Data Rate (b) Power Consumption Pie Chart (c) PSNR vs Buffer Current (d) Shmoo Plot }
    \label{fig:meas_results}
\end{figure*}

\subsection{Characterization and Calibration for Q Matrix} 
To implement the quantization block, a switched-capacitor-based method divides the output of 2D-DCT with elements of the quantization matrix. The quantization matrix (Q) is designed so that significant samples (top-left) are divided by small values, while insignificant samples (bottom-right) are divided by large values.


On-chip realization of the Q-matrix employs a capacitor bank, where charge from the 2D-DCT is directed to smaller capacitors for small divisions and larger capacitors for large divisions. However, process variations, parasitic capacitors from switches, and routing from the 2D-DCT block introduce changes in the capacitance values, impacting the element-wise division.
To address these variations, an intentional reduction in Q-matrix capacitance values is applied. Despite this, variations persist even after matching capacitors using parasitic-extracted simulations. To mitigate overall process variation, a calibration scheme adjusts the inverse Q-matrix to reconstruct the compressed image accurately.



Calibration involves sending 64 known (pre-determined) image frames to the dAJC IC, with a specific pixel value high and others zero as shown in Fig. \ref{fig:q_calibration}. If the measured output differs from the expected value, the inverse Q-matrix is adjusted to improve image quality. Fig. \ref{fig:q_calibration} illustrates that due to variations, the ideal Q-matrix is modified to Q', improving PSNR from 24.17dB to 33.21dB when using the modified inverse Q-matrix (Q''). This demonstrates a significant 10 dB improvement in PSNR through Q-matrix calibration.


\section{Measurement Results}
\subsection{Chip Micrograph and Test Setup}

\begin{figure}[htp]
   \centering
    \includegraphics[width=0.47\textwidth]{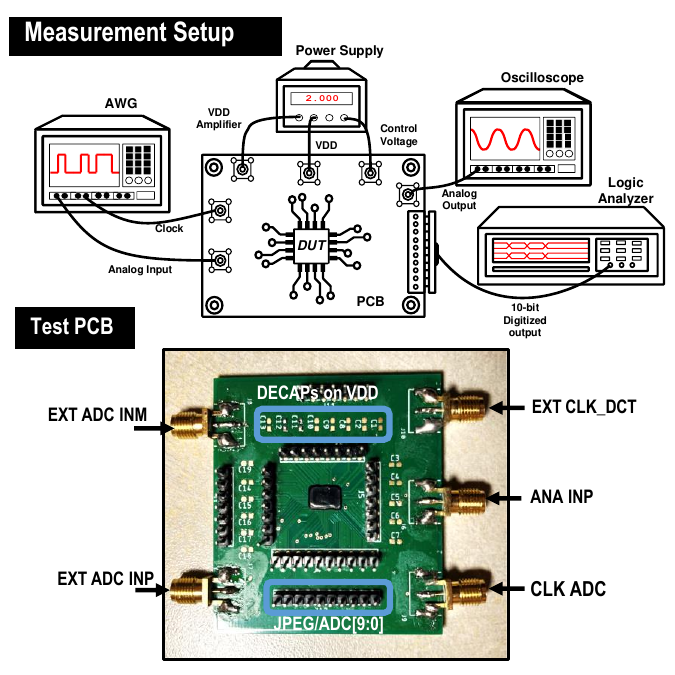}
    \caption{Measurement Setup and the test PCB}
    \label{Measurement_Setup}
    \vspace*{-0.2in}
\end{figure}

To validate the functionality of the dAJC IC and to demonstrate the operation of MJPEG video compression, the dAJC IC is fabricated in TSMC 65nm LP CMOS technology. Fig. \ref{ChipDie} shows the die micrograph and chip specifications of the fabricated IC. The design occupies a 1.25 mm $\times$ 2 mm area with an active area of 875um $\times$ 1639um. The measurement setup is shown in Fig. 
\ref{Measurement_Setup} along with the test PCB. The overall chip image has been divided into various critical blocks and highlights 
\begin{figure*}[htp]
   \centering
    \includegraphics[width=0.9\textwidth]{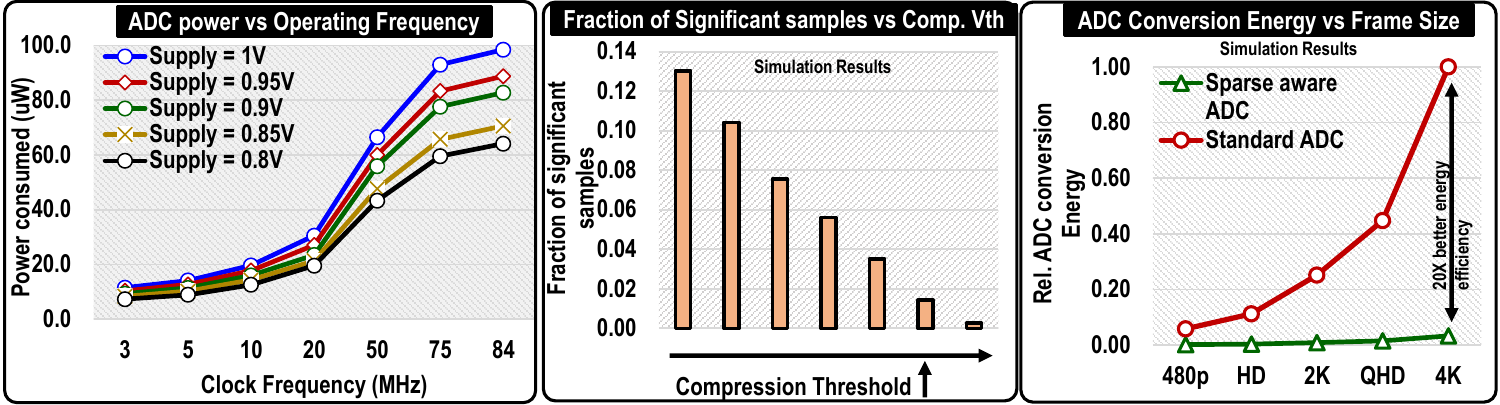}
    \caption{ADC Results: (a) Power Consumption vs Clock Frequency for various supply voltages (b) Simulation result showing the fraction of significant samples versus the compression threshold (c) ADC conversion Energy versus Frame Size showing 20X improvement in energy efficiency.  }
    \label{fig:adc_measurement_results}
    \vspace*{-0.1in}
\end{figure*}
to show the placement of the blocks and the flow of the signals. The hardware implementation of the Q-matrix has been shown on the IC, where the capacitors are arranged in an 8$\times$8 matrix form on the IC. For the measurement, the clock and the analog data samples are provided by the Arbitrary Waveform Generator(AWG)(enacts the working of an analog camera). There are separate power supplies for various blocks, to measure each block's power consumption. The compressed JPEG data is accessed in the analog form using the oscilloscope and then the waveform is used for further processing. The digitized compressed data is accessed by the 10-bit output pins JPEG/ADC[0:9]. The received compressed data is then processed with MATLAB to re-construct the image back and then compared with the original image to calculate the overall PSNR value. The overall performance of the system is measured using the PSNR metric. In the bottom part, Fig.\ref{Measurement_Setup} shows the test PCB for the IC, showing the various input and output pins. 

\subsection{Results}
Fig.\ref{fig:meas_results} shows the overall measurement results. From the power versus data rate plot it is observed that with the increase in the input data frequency, the total power consumption also increases. For a maximum input data rate of 5 MSps, the total power consumption is 2.02 mW.

From measurement data, it can be seen that the overall power consumption is dominated by the buffer power. The bandwidth, load capacitor, and slew requirements increase the buffer power.  For a sampling clock frequency of 500KHz and 10MHz, the PSNR is measured for a black-and-white image(exploits the maximum input range) for various bias currents of the buffer, which can be used to determine the minimum value of current the buffer requires so that the compression is correct and the image can be re-constructed back faithfully. It shows that for a higher clocking frequency, a higher bias current is required to keep the PSNR above 25dB, above which the image reconstructed back can be perceived easily by the human eye. So for a clock frequency of 500KHz and 10MHz, the bias current required is 26uA and 34uA respectively. Lastly, the operability area of the ASIC has been tested and measured using the shmoo plot. It shows that for a higher clock frequency, the minimum required VDD is higher. So for a clock frequency of 20MHz, the minimum required supply voltage is 0.95V, however, for a clock frequency of 2MHz, the minimum required supply voltage is 0.8V. But most failures shown in the shmoo are due to image quality degradation, although for these VDDs the circuit operates as required.

Results for ADC are shown in Fig.\ref{fig:adc_measurement_results}. The ADC power scales with the clock frequency, such that for the ADC to work at 84 MHz it consumes a power of 98.5 uW at a VDD of 1V. From simulation results, with the increase in the compression threshold, the fraction of significant samples reduces. As a result, the Sparsity-aware ADC outperforms the standard ADC, reducing the conversion energy by 20$\times$ as compared to the standard ADC (that is digitizing all the analog samples). The improved energy efficiency is achieved because the sparsity-aware ADC is digitizing only the significant samples that constitute less than 5\% of the total samples. Fig.\ref{fig:adc_measurement_results}(c). shows the relative energy efficiency comparing Sparsity-Aware ADC to Standard ADC for various frame sizes (which controls the overall incoming data rate). 
\section{Comparison with related works}
In this section, the design is compared against its digital implementation and the state-of-art implementations.
\begin{figure}[htp]
   \centering
    \includegraphics[width=0.5\textwidth]{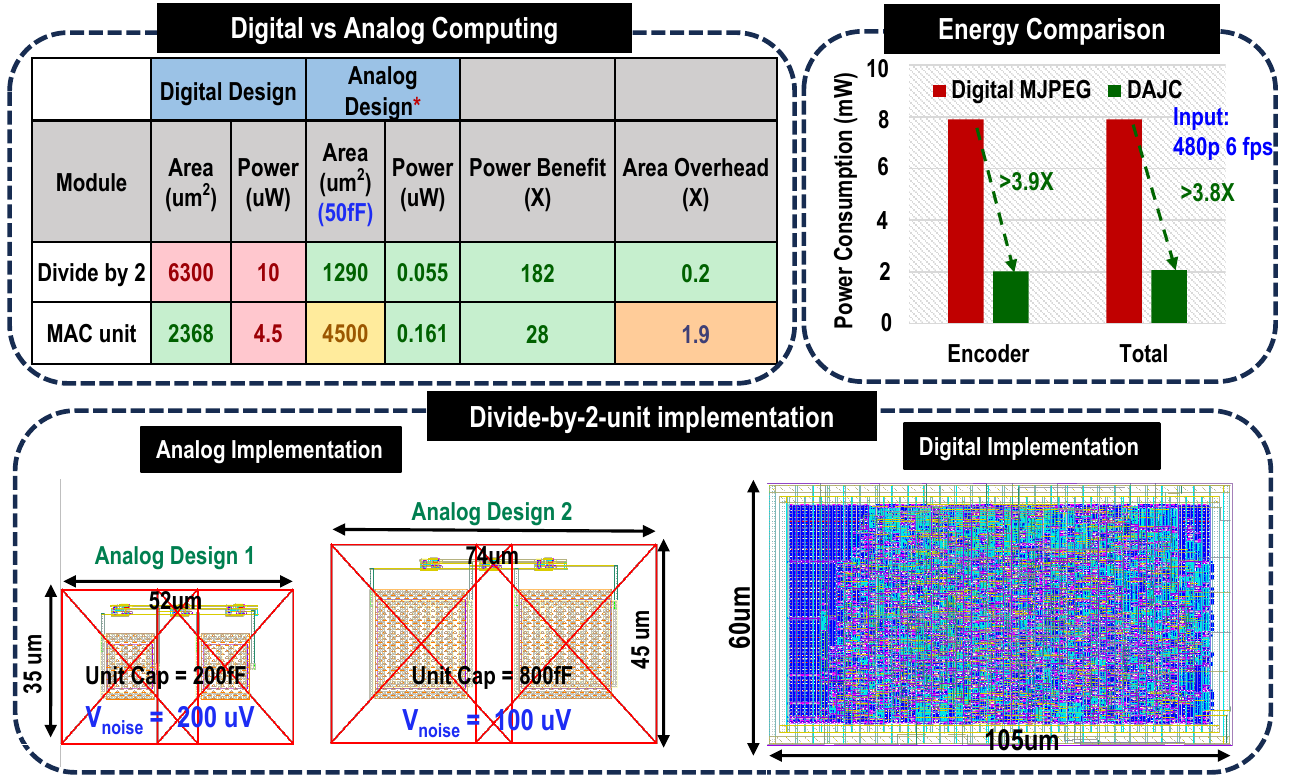}
    \caption{Comparison of the Current implementation with its Digital version in terms of Power and Area.}
    \label{fig:analog_vs_digital}
    \vspace*{-0.2in}
\end{figure}

\begin{figure*}[htp]
   \centering
    \includegraphics[width=0.95\textwidth]{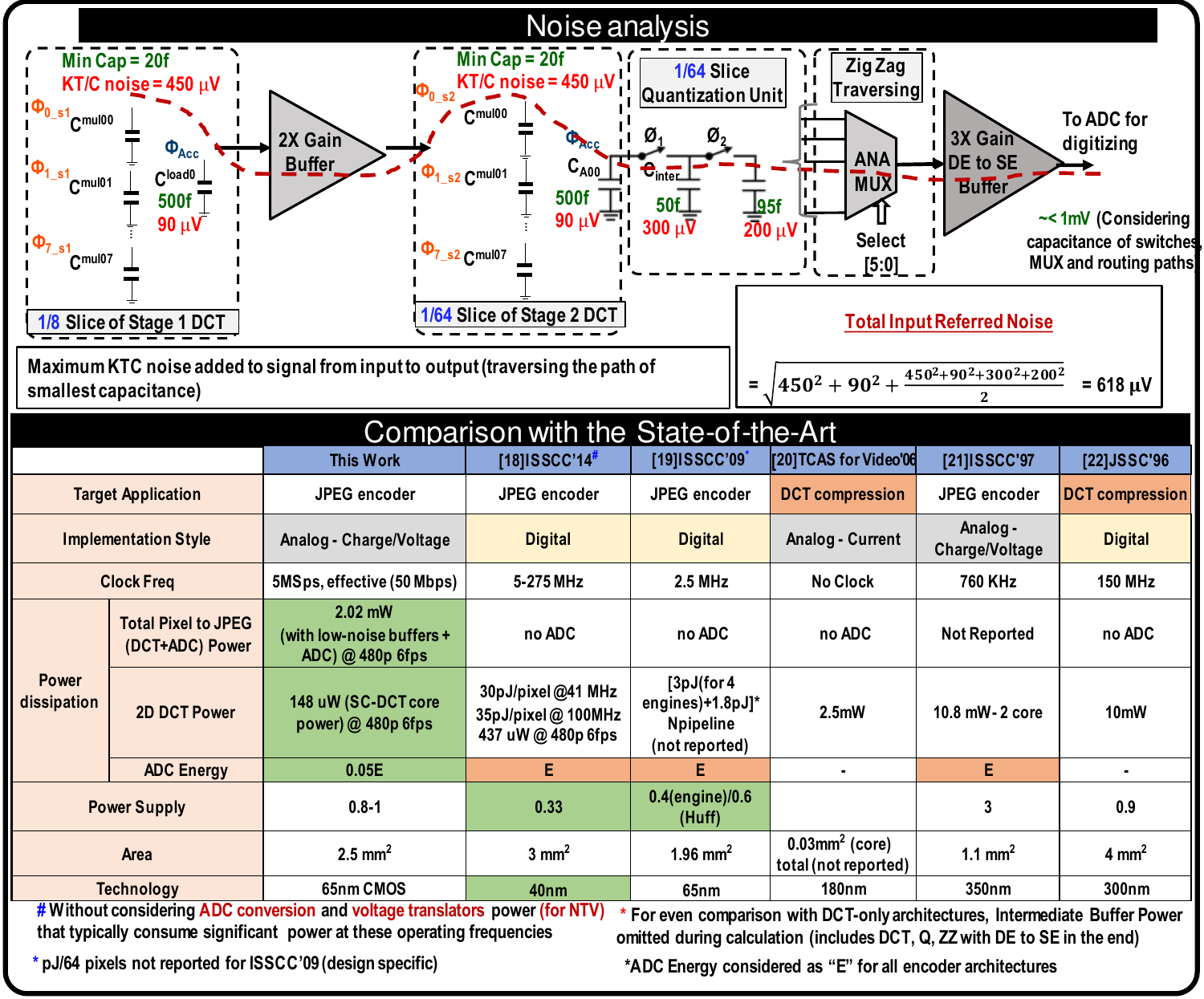}
    \caption{Noise analysis and Comparison table with the state-of-the-art implementations.}
    \label{fig:noise_and_comparison_table}
    \vspace*{-0.25in}
\end{figure*}
\subsection{Digital vs Analog Implementation}

The reason for the selection of the analog architecture can be understood by comparing it with the digital implementation. Fig.\ref{fig:analog_vs_digital} shows the analog vs digital comparison. The digital architecture is digitally synthesized and encoding power is calculated using spice simulations. A 3.9$\times$ overall power benefit is achieved by using an analog implementation of the JPEG compression block compared to its digital implementation as shown in the energy comparison plot. Comparison has also been done for the basic building blocks for the compression unit, which are the division unit cell and the MAC unit cell. The overall power benefit of analog implementation against digital implementation for implementing a divide by 2 unit and MAC unit is 182$\times$ and 28$\times$ respectively. The area benefit is 2$\times$ for implementing a MAC unit. Considering the area for implementing a divide by 2 unit, using a large capacitor to limit KTC noise, the overall area of analog implementation is approximately half of its digital implementation. Hence analog processing before the ADC conversion helps save power and eliminates the need to digitize all the sensed analog samples for processing, lowering the Front-End ADC conversion energy. 

\subsection{Comparison against state of the art}
In this section, dAJC is compared against the previous IC implementations\cite{reynders201427},\cite{yu_pu_ultra-low-energyframe_2009},\cite{pankaala_analog_2006},\cite{kawahito1997compressed},\cite{kuroda_09-v_1996} as shown in Fig.\ref{fig:noise_and_comparison_table}. The Switched-capacitor-based low-noise 2D DCT implementation consumes 148uW power, which is $\sim12X$ lower than that of other reported works\cite{pankaala_analog_2006}. This can be attributed to the absence of any active circuitry, like an operational amplifier (OPAMP) for multiplying and accumulating MAC results. The operations are performed carefully using passive circuits, i.e., switches, and appropriately-sized capacitances. The introduction of sparsity-aware ADC assists in reducing the total ADC energy as it is activated only for significant samples while being powered down for insignificant ones. Statistically, the number of significant samples after JPEG encoding is less than 5\%, which consequently reduces the ADC energy by greater than 95\%.  

\section{Discussions}
This section provides a deeper analysis of the capacitor value selection and small design considerations taken for the proper functioning of the IC as shown in Fig.\ref{fig:noise_and_comparison_table}. Since all the operations are switched-capacitor-based, the calculation for the overall KT/C noise in the system \cite{7182091},\cite{8702129}  is very crucial. The overall noise is dominated by the smallest capacitor used at each stage. Apart from this the addition of the buffers in between the Stage 1 DCT and Stage 2 DCT also alleviates the inferred KTC noise at the input. From the calculation, the overall input referred to KTC noise is $\sim$ 600uV. Now while selecting the minimum capacitor value, a minimum capacitor value of 20fF has been chosen such that the overall area can be reduced and the effect of the parasitics does not affect the overall result. Further, going to lower technology nodes with high-density capacitors can significantly lower the overall area of the chip and enable further low-power implementation.

\section{Conclusion}

In conclusion, the first end-to-end implementation of MJPEG compression in the Mixed Signal Domain has been realized on 65nm CMOS technology. This endeavor aimed to transition compression to the analog domain to diminish digital computation power consumption, a goal successfully demonstrated in this study. Moreover, by relocating compression to the analog domain, processing can be conducted in-sensor, enabling signal analysis directly at the sensor without the need to transmit data to the main processing hub for information extraction. Consequently, by integrating MJPEG compression at or near the sensor, the integrated circuit achieves $\sim4$ times lower power consumption compared to its digital implementations. Additionally, a 20-fold reduction in ADC conversion energy is attained through the utilization of sparsity-aware ADC. Furthermore, compressing the signal in the analog domain diminishes the overall power expended for communication and transmission.   






%
\bibliographystyle{unsrt}
{
    \bibliography{arxiv.bib}
}

\vskip -1\baselineskip plus -1fil

\begin{IEEEbiography}[{\includegraphics[width=1in,height=1.25in,clip,keepaspectratio]{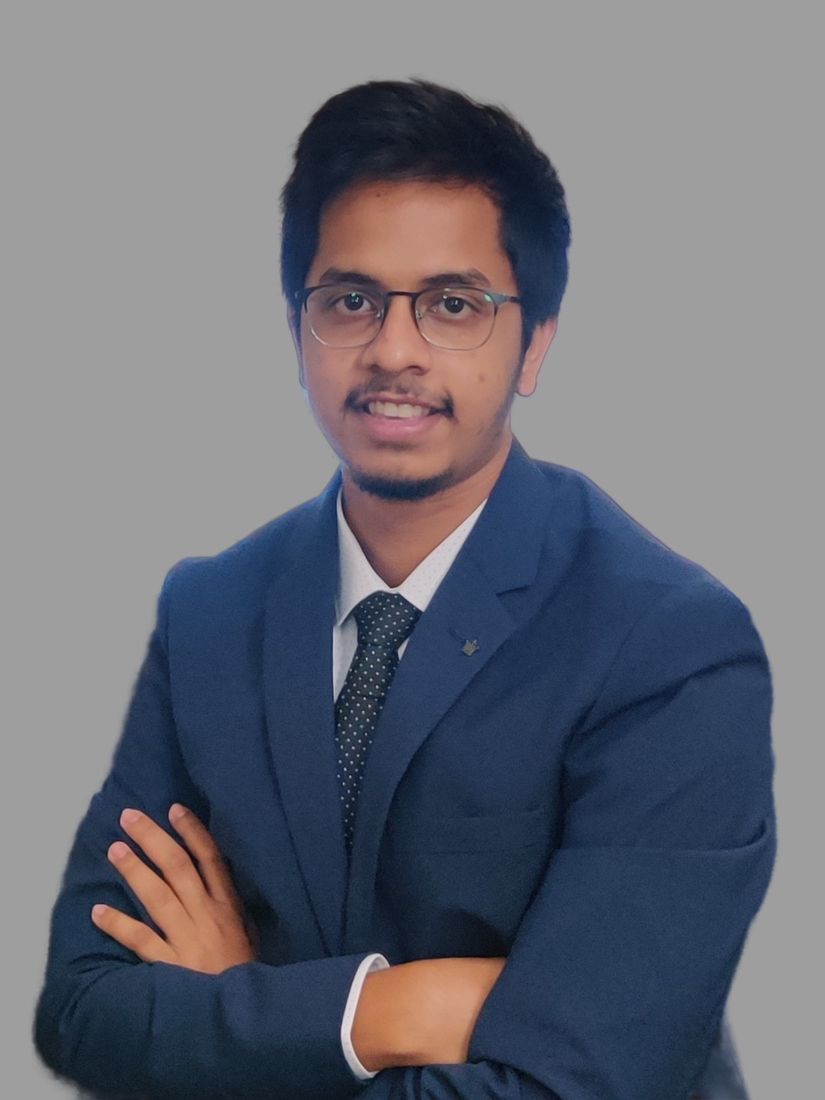}}]
{Gourab Barik}
 (S`21) (Graduate Student Member, IEEE) received a bachelor’s degree in Electronics and Telecommunication Engineering from the Indian Institute of Engineering Science and Technology, Shibpur, India, in 2019. He is currently pursuing a Ph.D. degree in Electrical and Computer engineering at Purdue University, West Lafayette, IN, USA, working with Prof. Shreyas Sen. 
 
Before joining Purdue, he worked as an Analog Design Engineer at Tagore Technology, India working on GaN and CMOS-based PMICs. His research interests include mixed-signal IC design for low-power and high-speed sensors for Bio-Medical applications.
 
 He has served as a reviewer for multiple reputed conferences, including the IEEE International Conference on VLSI Design (VLSID) and IEEE BioCAS.
\end{IEEEbiography}

\vskip -1\baselineskip plus -1fil

\begin{IEEEbiography}[{\includegraphics[width=1in,height=1.25in,clip,keepaspectratio]{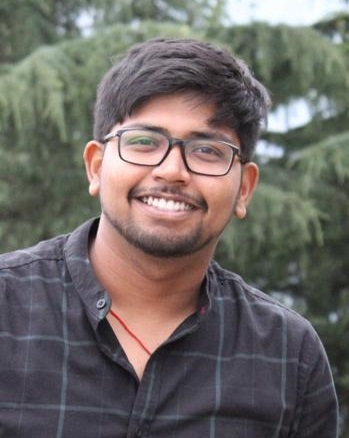}}]
{Gaurav Kumar K}
 (S`19) (Graduate Student Member, IEEE) received a bachelor’s degree in Electronics and Telecommunication Engineering from Jadavpur University, Kolkata, India, in 2017. He is currently pursuing a Ph.D. degree in electrical and computer engineering with Purdue University, West Lafayette, IN, USA.
 
Before joining Purdue, he worked as a Network Experience manager in the telecom industry at Bharti Airtel, India. He has interned with the Advanced R\&D Discovery Team, Texas Instruments (TI) Kilby Labs, USA, during the summer of 2021, where he worked on designing RX front-end circuits for high-speed electrical links. His research interests include mixed-signal IC design for low-power and high-speed sensors and wireline links.
 
 He has served as a Primary and Secondary Reviewer for multiple reputed conferences, including the IEEE International Conference on VLSI Design (VLSID) and the IEEE Custom Integrated Circuit Conference (CICC).
\end{IEEEbiography}

\vskip -1\baselineskip plus -1fil

\begin{IEEEbiography}[{\includegraphics[width=1in,height=1.25in,clip,keepaspectratio]{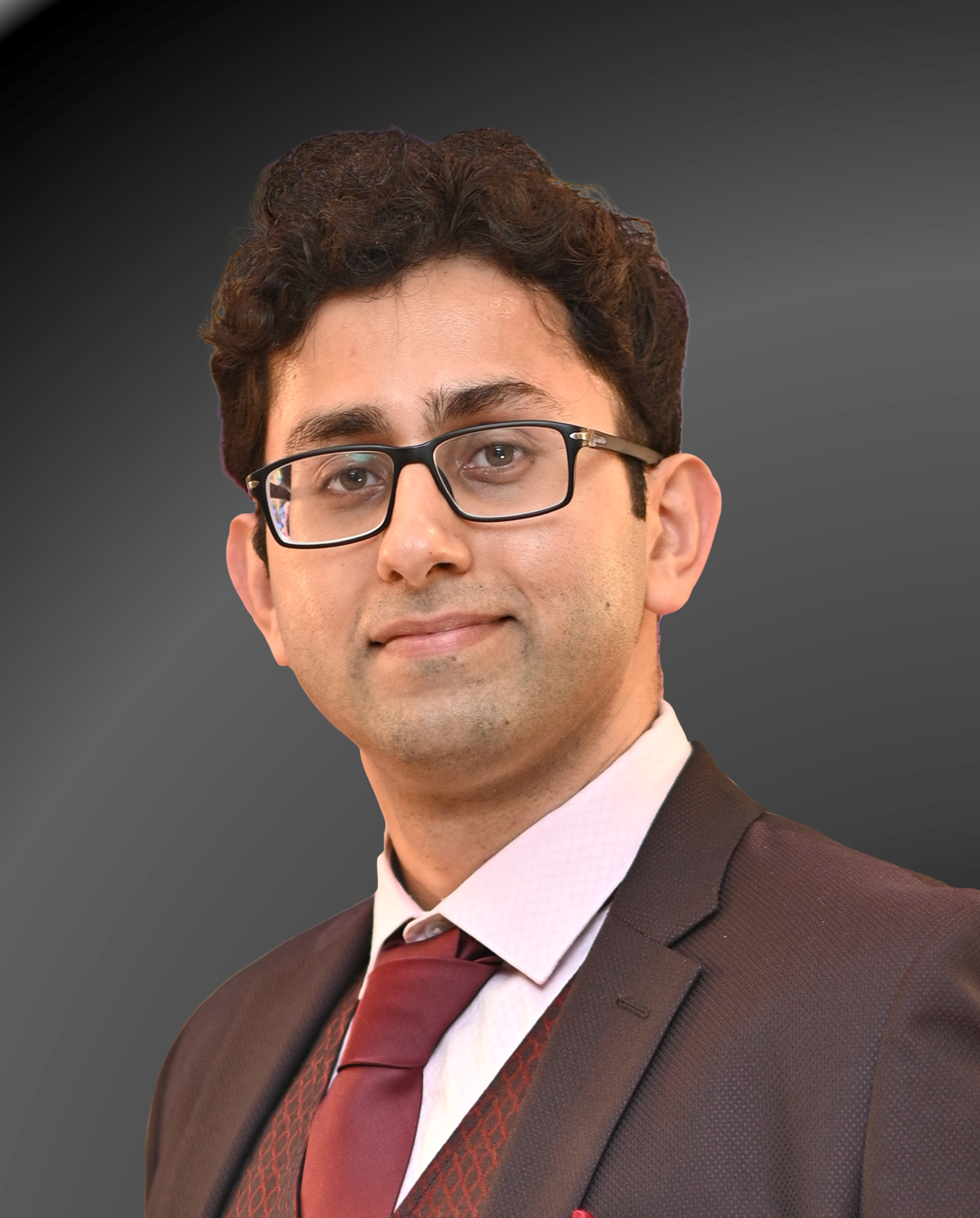}}]
{Baibhab Chatterjee}
 (S`10, M`22) received his Ph.D. from the Elmore Family School of Electrical Engineering, Purdue University, West Lafayette, IN, USA in 2022. He is currently an Assistant Professor in the department of ECE, University of Florida, Gainesville, FL, USA. His industry experience includes two years as a Digital Design Engineer/Senior Digital Design Engineer with Intel, Bengaluru, India, and one year as a Research and Development Engineer with Tejas Networks, Bengaluru, India. He was a Quantum Hardware Design Intern with IBM T.J. Watson Research Center, NY, USA during 2020-2021, where he worked on ultra-low power quantum receiver front-ends.
 
 Dr. Chatterjee received the University Gold Medal from NIT, Durgapur, India, in 2011, the Institute Silver Medal from IIT Bombay in 2015, the Andrews Fellowship at Purdue University during 2017-2019, the HOST 2018 Best Student Poster Award (3rd), the CICC 2019 Best Paper Award (overall), the RFIC/IMS 2020 3MT Award (audience choice) and the Bilsland Dissertation Fellowship at Purdue University during 2021-2022. His research interests include low-power analog, RF, and mixed-signal circuit design for next-generation biomedical and quantum applications.
\end{IEEEbiography}

\vskip -1\baselineskip plus -1fil

\begin{IEEEbiography}[{\includegraphics[width=1in,height=1.25in,clip,keepaspectratio]{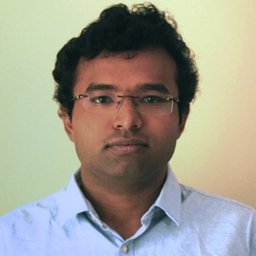}}]
{Shovan Maity}
(S’18) received the B.E. degree from Jadavpur University, Kolkata, India, in 2012, the M.Tech. degree in electrical engineering from IIT Bombay, Mumbai, India, in 2014, and the Ph.D. degree in electrical engineering from Purdue University, West Lafayette, IN, USA, in 2019. He worked as an Analog Design Engineer with Intel, Bengaluru, India, from 2014 to 2016 and as a Senior Circuit Design Engineer with Qualcomm, San Diego, CA, USA from 2019 to 2021. He is currently working as a Staff Electrical Engineer at Ixana, West Lafayette, IN, USA. His research interests lie in the area of mixed-signal circuits and systems for the Internet of Things, and biomedical and security applications. Dr. Maity received the Institute Silver Medal from IIT Bombay in 2014, the Purdue ECE Fellowship from 2018 to 2019, the IEEE HOST Best Student Paper Award in 2017, and the CICC 2019 Best Paper Award.
\end{IEEEbiography}

\vskip -1\baselineskip plus -1fil

\begin{IEEEbiography}[{\includegraphics[width=1in,height=1.25in,clip,keepaspectratio]{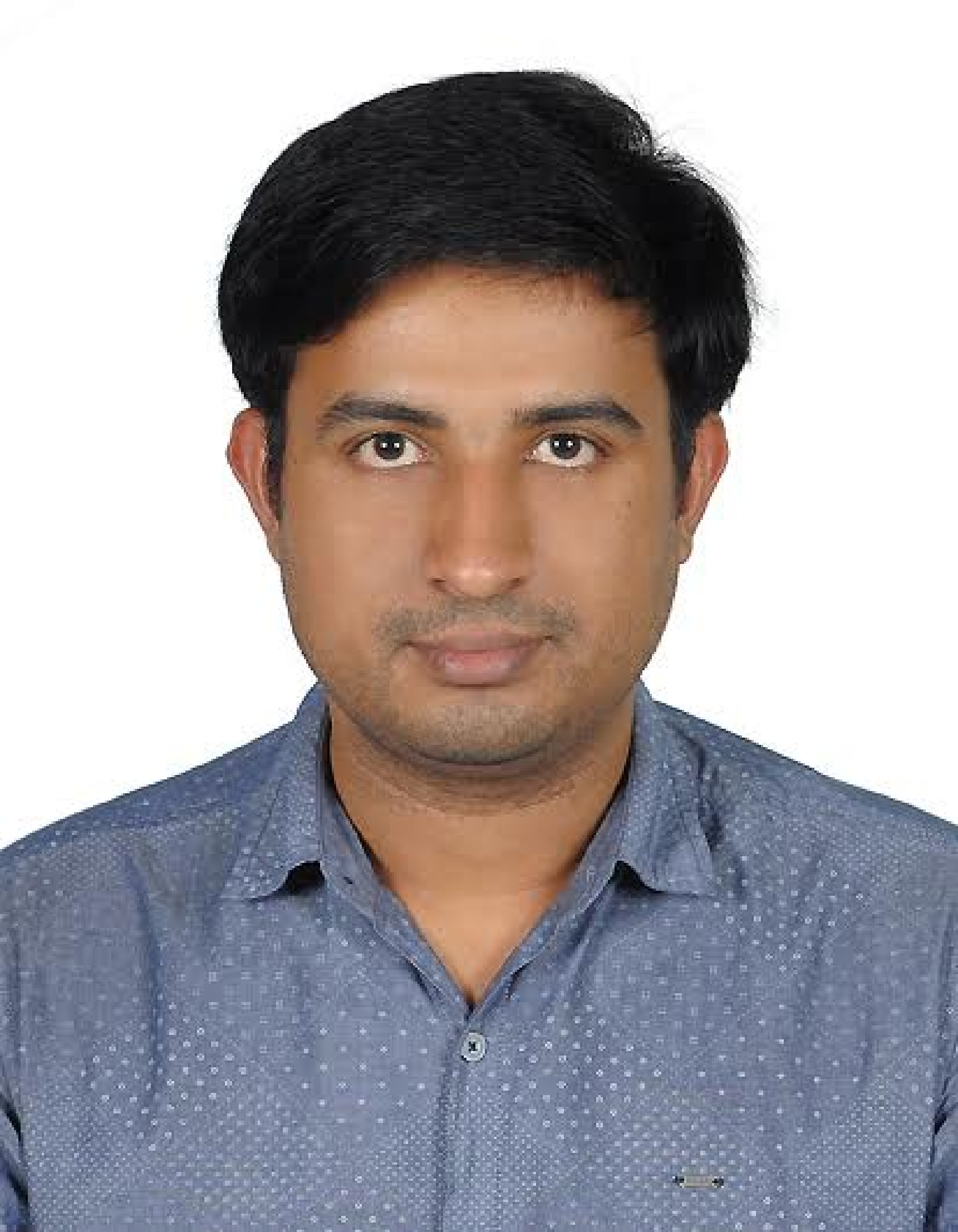}}]
{Sumon Bose}
(S'18) received a Bachelor of Engineering in Electronics and Telecommunication Engineering from Jadavpur University, Kolkata in 2012 and a Ph.D. degree from Nanyang Technological University, Singapore in 2021. After graduation, he worked in several companies like Cypress Semiconductor, xSi Semiconductor, and Texas Instruments where he worked on PMIC.

Currently, he is an Analog design engineer at Quasistatics Inc. His research interests include low-power analog IC and Machine Learning-based Hardware Design.
\end{IEEEbiography}

\vskip -1\baselineskip plus -1fil

\begin{IEEEbiography}[{\includegraphics[width=1in,height=1.25in,clip,keepaspectratio]{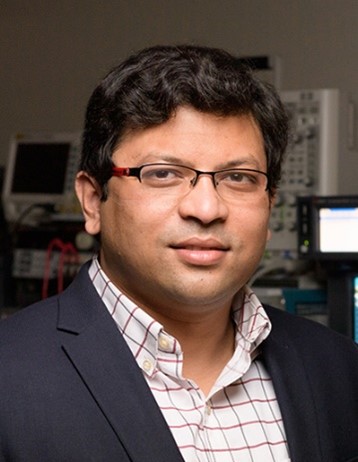}}]
{Shreyas Sen} (S`06-M`11-SM`17) (Senior Member, IEEE) is an Elmore Associate Professor of ECE \& BME, Purdue University. His current research interests span mixed-signal circuits/systems and electromagnetics for the Internet of Bodies (IoB) and Hardware Security. He has authored/co-authored 3 book chapters, over 200 journal and conference paper and has 25 patents granted/pending. Dr. Sen serves as the Director of the Center for Internet of Bodies (C-IoB) at Purdue.
Dr. Sen is the inventor of the Electro-Quasistatic Human Body Communication (EQS-HBC), or Body as a Wire technology, for which, he is the recipient of the MIT Technology Review top-10 Indian Inventor Worldwide under 35 (MIT TR35 India) Award in 2018 and Georgia Tech 40 Under 40 Award in 2022. To commercialize this invention Dr. Sen founded Ixana and serves as the Chairman and CTO and led Ixana to awards such as 2x CES Innovation Award 2024, EE Times Silicon 100, Indiana Startup of the Year Mira Award 2023. His work has been covered by 250+ news releases worldwide, invited appearance on TEDx Indianapolis, NASDAQ live Trade Talks at CES 2023, Indian National Television CNBC TV18 Young Turks Program, NPR subsidiary Lakeshore Public Radio and the CyberWire podcast. Dr. Sen is a recipient of the NSF CAREER Award 2020, AFOSR Young Investigator Award 2016, NSF CISE CRII Award 2017, Intel Outstanding Researcher Award 2020, Google Faculty Research Award 2017, Purdue CoE Early Career Research Award 2021, Intel Labs Quality Award 2012 for industrywide impact on USB-C type, Intel Ph.D. Fellowship 2010, IEEE Microwave Fellowship 2008, GSRC Margarida Jacome Best Research Award 2007, and nine best paper awards including IEEE CICC 2019, 2021 and in IEEE HOST 2017-2020, for four consecutive years. Dr. Sen's work was chosen as one of the top-10 papers in the Hardware Security field (TopPicks 2019). He serves/has served as an Associate Editor for IEEE Solid-State Circuits Letters (SSC-L), Nature Scientific Reports, Frontiers in Electronics, IEEE Design \& Test, Executive Committee member of IEEE Central Indiana Section and Technical Program Committee member of TPC member of ISSCC, CICC, DAC, CCS, IMS, DATE, ISLPED, ICCAD, ITC, and VLSI Design. Dr. Sen is a Senior Member of IEEE.
\end{IEEEbiography}

\end{document}